\documentclass[12pt]{emulateapj} 
\usepackage{natbib}
\usepackage{graphicx}
\usepackage{color}
\usepackage{amsmath}

\makeatletter
\newcommand*{\rom}[1]{\expandafter\@slowromancap\romannumeral #1@}
\makeatother

\begin{document}                          
\title{Numerical convergence in smoothed particle hydrodynamics}

\author{
Qirong Zhu$^{1,2}$, Lars Hernquist$^{3}$, Yuexing Li$^{1,2}$,
}

\affil{$^1$Department of Astronomy \& Astrophysics, The Pennsylvania State University, 
525 Davey Lab, University Park, PA 16802, USA}
\affil{$^2$Institute for Gravitation and the Cosmos, The Pennsylvania State University, University Park, PA 16802, USA}
\affil{$^2$Harvard-Smithsonian Center for Astrophysics, 60 Garden Street, Cambridge, MA 02138, USA}

\begin{abstract}

We study the convergence properties of smoothed particle hydrodynamics
(SPH) using numerical tests and simple analytic considerations.  Our
analysis shows that formal numerical convergence is possible in SPH
only in the joint limit $N \rightarrow \infty$, $h \rightarrow 0$, and
$N_{nb} \rightarrow \infty$, where $N$ is the total number of
particles, $h$ is the smoothing length, and $N_{nb}$ is the number of
neighbor particles within the smoothing volume used to compute
smoothed estimates.  Previous work has generally assumed that the
conditions $N \rightarrow \infty$ and $h \rightarrow 0$ are sufficient
to achieve convergence, while holding $N_{nb}$ fixed.  We demonstrate
that if $N_{nb}$ is held fixed as the resolution is increased, there
will be a residual source of error that does not vanish as $N
\rightarrow \infty$ and $h \rightarrow 0$.  Formal numerical
convergence in SPH is possible only if $N_{nb}$ is increased
systematically as the resolution is improved.  Using analytic
arguments, we derive an optimal compromise scaling for $N_{nb}$ by
requiring that this source of error balance that present in the
smoothing procedure.  For typical choices of the smoothing kernel, we
find $N_{nb} \propto N^{0.5}$.  This means that if SPH is to be used
as a numerically convergent method, the required computational
cost does not scale with particle
number as $O(N)$, but rather as $O(N^{1+\delta})$, where $\delta
\approx 0.5$, with a weak dependence on the form of the smoothing
kernel.

\end{abstract}

%\keywords{methods: numerical -- hydrodynamics}

\section{Introduction}

Smoothed particle hydrodynamics (SPH) has become a popular tool for
studying astrophysical flows since its introduction in the mid-1970s
by \cite{Lucy1977} and \cite{Monaghan1977}.  Instead of solving the
equations of hydrodynamics on a mesh, SPH operates by associating
fluid elements with particles that characterize the flow.  Local
properties are estimated by ``smoothing'' the attributes of particles
(neighbors) near a given point.  Typically, the particles carry a
fixed mass (unless converted to another form, such as collisionless
stars), and the smoothing is done in a spherically symmetric manner
using an interpolation function known as the smoothing kernel.
Because it is gridless, SPH is naturally spatially adaptive if the
scale of the smoothing, set by the smoothing length $h$, varies as the
particles move with the flow.  If individual particle timesteps are
employed, the scheme will also be temporally adaptive.

SPH has been used with success especially in investigations of the
formation and evolution of galaxies and large scale structure, where
spatial and temporal adaptivity are essential because of the large
range in scales present.  Along with the successes has come an
understanding of various limitations with SPH.  Examples include: an
artificial clumping instability for particular choices of the
smoothing kernel \citep{Schussler1981}; a lack of energy or entropy
conservation for various forms of the equations of motion when the
smoothing lengths vary spatially \citep{Hernquist1993}; and a poor
handling of instabilities at shearing interfaces owing to artificial
surface tension effects \citep{Agertz2007}.  Many of the problems have
already been overcome through a variety of improvements to the
original formulation of SPH, such as the introduction of generalized
smoothing kernels \citep{Dehnen2012}, fully conservative equations of
motion based on variational principles \citep{Springel2002}, and
various methods to eliminate artificial surface tension effects
\citep{Price2008,Read2009, Hopkins2013, Saitoh2013}.

However, some issues remain unresolved, which has led to numerous
misconceptions within the community of users of SPH.  For example,
because densities in SPH are calculated using smoothed estimates,
rather than by solving the mass continuity equation, the flow of the
fluid on small scales is not described correctly, as emphasized by
\cite{Vogelsberger2012}.  One claimed advantage of SPH over grid
based methods is that the previous history of fluid elements can be
determined straightforwardly because individual particles retain their
identity over time.  With mesh codes, this can be accomplished only by
introducing tracer particles into the flow so that the evolution of
the density field can be reconstructed \citep{Genel2013}.  In fact,
this ``advantage'' of SPH owes to the fact that this method does not
solve the equations of motion correctly on the smoothing scale.

Here, we focus on another issue with SPH that has received little
attention in the literature, but for which confusion abounds: the
numerical convergence of this technique.  A commonly used approach for
making SPH spatially adaptive is to allow the smoothing length of each
particle to decrease or increase depending on the local density of
particles.  One possible choice is to scale the smoothing lengths in
proportion to the mean interparticle separation according to $h
\propto N^{-1/3}$ \citep{Hernquist1989}.  A benefit to this scaling is
that the number of particles used to compute smoothed estimates
(particles neighboring a given point), $N_{nb}$, is roughly constant,
optimizing the efficiency of the method from one region to another.
But, this choice is somewhat arbitrary and leads to difficulties that
are not widely recognized.

In SPH, local quantities are estimated by smoothing continuous fluid
variables with an interpolation kernel.  The resulting convolutions
are evaluated numerically by approximating the integrals with discrete
sums.  These sums are calculated over $N_{nb}$ fluid elements
(particles) near a given point.  For the scaling noted above $N_{nb}
\approx$ constant, and is typically limited to relatively small values
$N_{nb} \approx 30-100$ to maximize spatial resolution.  However, the
approximation of the convolution integrals by discrete sums entails an
error that depends directly on $N_{nb}$ and not $N$ or $h$.
Therefore, as the resolution is improved by increasing $N$ and
reducing $h$, numerical convergence is not possible because the error
in the discrete estimates does not vanish if $N_{nb}$ is held
constant.

Systematic studies of the convergence properties of SPH are lacking,
but evidence that this source of error exists is present in the
literature.  For example, in his review of SPH, \cite{Springel2010}
performed a series of tests to empirically determine the convergence
rate of SPH.  The behavior we suggest can be seen clearly in the
Gresho vortex problem, as in Fig. 6 of \cite{Springel2010}.  Here,
various runs were performed at improving resolution by increasing the
total number of particles, $N$, and reducing the smoothing lengths,
$h$, but holding the number of neighbors, $N_{nb}$ fixed.  The solid
curves in this figure compare the error in each case to a known
analytic solution.  For small particle numbers, increasing $N$
decreases the error, but eventually the error saturates and plateaus
at value that depends in detail on the choice of artificial viscosity.
However, if the runs with varying $N$ are compared to the highest
resolution versions (dashed curves), the error shows a monotonically
decreasing trend with $N$, for all choices of the artificial viscosity.
This behavior can be explained only by the presence of a residual
source of error that does not depend directly on $N$ and $h$.  In what
follows, we show that this error arises in the discrete sums used to
approximate the local smoothing convolutions and that it depends
directly only on the number of neighbors in these sums, $N_{nb}$.

For a random distribution of particles, this residual error
essentially arises from shot noise in the discrete estimates and is
expected to scale as $\propto N_{nb}^{-0.5}$.  However, as emphasized
by \cite{Monaghan1992}, SPH particles are usually not randomly
distributed but are instead arranged in a ``quasi-regular'' pattern
because of local forces between them.  An analysis of the error in
this situation suggests that the noise should scale with $N_{nb}$ as
$\sim (d-1) N_{nb}^{-1} \log N_{nb}$, where $d$ is the number of
dimensions.  (Note that to this order of accuracy, the error estimate
vanishes for $d=1$, meaning that 1-dimensional tests cannot be used to
gauge the numerical convergence of SPH for $d>1$, contrary to common
belief.)

Using simple numerical tests, we verify that the expected $\sim
N_{nb}^{-1} \log N_{nb}$ behavior is a reasonable characterization of
the discreteness error for multi-dimensional simulations with SPH.
Therefore, the discreteness error,
error$_d$, can be bounded by the estimates $N_{nb}^{-1} < $
error$_d$ $< N_{nb}^{-0.5}$.  Using this scaling, we suggest an
optimal compromise between accuracy and efficiency by requiring that
this error decline at the same rate as the error associated with the
smoothing procedure itself, which directly involves only the smoothing
lengths, $h$.  In detail, the outcome depends on the form of the
smoothing kernel, but for typical choices we find that the number of
neighbors in the smoothed estimates should increase with the total
particle number as $N_{nb} \propto N^{0.5}$.  In this case, the
smoothing lengths would scale as $h \propto N^{-1/6}$, rather than with
the average separation between particles as $h \propto N^{-1/3}$, in
which case $h$ represents the geometric mean of the size of the system
and the average interparticle separation.

Our empirical tests verify that numerical convergence can indeed be
obtained with SPH, if the neighbor number grows systematically with
improved resolution.  Of course, this comes with the consequence of
increased computational expense relative to holding $N_{nb}$ fixed,
and implies an unfavorable cpu scaling of $O(N^{3/2})$ compared to
grid codes which are typically $O(N)$.  This should not be surprising,
since SPH is a Monte Carlo-like algorithm and it is well-known in
other contexts that, while being flexible and general, Monte Carlo
methods typically have unfavorable convergence properties relative to
other more specialized techniques.  Ultimately, applications with SPH
must therefore choose between accuracy and computational efficiency.
What is clear, however, is that if SPH is used with constant $N_{nb}$,
numerical convergence is not possible.

We emphasize that this issue with the convergence of SPH is well-known
in some disciplines (e.g. \citealt{Robinson2012}).  However, the specific
problem we analyze here has been virtually
ignored in discussions of the use of SPH in
applications to cosmology and galaxy formation.

\section{Consistency and Self-consistency}

In SPH, densities are estimated from the particles by smoothing.  This
involves convolving a continuous field quantity, $A ({\mathbf{r}})$,
with a smoothing function $W(\mathbf{r}, h)$ through
\begin{equation}
A_s (\mathbf{r})  = \int A ({\mathbf{r}}) W(\mathbf{r}-\mathbf{r^{\prime}}, h) d\mathbf{r}^{\prime},
\end{equation}
where $A(\mathbf{r})$ and $A_s(\mathbf{r})$ denote the true field and
its smoothed version.  The smoothing length $h$ represents a
characteristic width of the kernel over which the desired quantity is
spread. The smoothing kernel is normalized to unity and asymptotically
approaches a Dirac $\delta$-function when $h \rightarrow 0$. This can be
achieved with a sufficiently large number of particles to describe the
continuous system as $N \rightarrow \infty$.

Assuming that the volume within the smoothing kernel is sufficiently
sampled with these discrete points, we can further approximate the
above integral by discrete summations. The ``volume"
element $\Delta \mathbf{r}^{\prime}$ is estimated from $m_b/\rho_b$
where $m_b$ and $\rho_b$ are the mass and density of particle $b$ and
we have
\begin{equation} \label{eq:sph_density}
A_d(\mathbf{r}) = \sum_b A_b \frac{m_b}{\rho_b} W(\mathbf{r-r}_b, h).
\end{equation}

With this approximation, we immediately obtain the density estimate in
its discrete form $\rho_d(\mathbf{r})$ as used in SPH according to:
\begin{equation} \label{eq:sphdensity}
\rho_d(\mathbf{r}) = \sum_b{m_b} W(\mathbf{r-r}_b, h),
\end{equation}
where the sum is over all the particles within the volume
element centered at $\mathbf{r}$.  A smoothing function with compact
support is usually used to limit the ``volume" to finite extent in
order to minimize computation time. Equation \eqref{eq:sphdensity}
now operates on a finite number $N_{nb}$ of particles near
a given point.  In order to satisfy the
consistency of this step, which is to approach the continuous limit
with such finite summations, the condition $N_{nb} \rightarrow \infty$
is required.

Moreover, in applications with finite $N$ and
$N_{nb}$, there is a lack of self-consistency based on the above
approach.  If we would represent a scalar field $A(\mathbf{r})$ with a
constant value, say $1$, the following condition
\begin{equation} \label{eq:self-consistency}
1 = \sum_b \frac{m_b}{\rho_b} W(\mathbf{r-r}_b, h),
\end{equation}
is generally not satisfied with the density field estimated from \eqref{eq:sphdensity}.

This lack of self-consistency is most extreme where large
density gradients are present, e.g. between two phases of flow
\citep{Read2009} or close to a boundary \citep{Liu2006}. In other
words, the volume estimate in SPH from equation \eqref{eq:sphdensity}
with a finite $N_{nb}$ is not an accurate partitioning of space. In order
to satisfy the condition of consistency and self-consistency, the
following conditions
\begin{equation}
N \rightarrow \infty ,  h  \rightarrow 0,  N_{nb} \rightarrow \infty
\end{equation}
indeed should be met if no other fixes such as normalization are applied. 

Next we discuss the consistency of the Euler equations that SPH
actually solves.  The original Euler equations in Lagrangian form
describe the evolution of density, momentum, and internal energy
according to:
\begin{equation}
\frac{d\rho}{dt}  = -\rho\nabla\cdot\mathbf{v} 
\end{equation}

\begin{equation}
\frac{d\mathbf{v}}{dt}  = -\frac{\nabla P}{\rho}
\end{equation}

\begin{equation}
\frac{du}{dt}  = -\frac{P}{\rho} \nabla \cdot \mathbf{v}
\end{equation}
where $\rho$, $\mathbf{v}$ and $u$ represent the density, velocity, and
internal energy per unit mass.  The equation of state

\begin{equation}
P = u (\gamma - 1) \rho
\end{equation}
is a closure equation for the above system, where $\gamma$ is the adiabatic index of the gas. 
Following the derivation in \cite{Springel2002}, the equations of motion
can be discretized as
\begin{equation}\label{eq:sph_cont}
\frac{d\rho_a}{dt} = f_a \sum_b m_b (\mathbf{v}_a  - \mathbf{v}_b) \cdot \nabla_a W(h_a)
\end{equation}

\begin{equation}\label{eq:sph_eom}
\frac{d\mathbf{v}_{a}}{dt} = -\sum_{b} m_b(\frac{f_a P_a}{\rho_a^2} \nabla_a W_{ab}(h_a) + \frac{f_b P_b}{\rho_b^2} \nabla_a W_{ab}(h_b))
\end{equation}

\begin{equation}\label{eq:sph_energy}
\frac{du_a}{dt} = f_a \frac{P_b}{\rho_b^2} \sum_b m_a (\mathbf{v}_a  - \mathbf{v}_b) \cdot \nabla_a W(h_a) ,
\end{equation}
where the factors $f_a$ and $f_b$ depend on derivatives of the density
with respect to the smoothing lengths.
In practice instead of $u$, we can also use a variable $A(s)$ from $P = A(s)\rho^\gamma$ to solve the energy equation as in \cite{Springel2002}. These two approaches are equivalent in principle. 

\cite{Read2009} have calculated the errors in the continuity and
momentum equations with the above SPH formulation assuming a finite $N_{nb}$
(ignoring the factors $f_a$, which spoils the fully conservative
nature of the equations) as
\begin{equation}
\frac{d\rho_a}{dt}  \approx - \rho_a (\mathbf{R}_a \nabla_{a}) \cdot \mathbf{v}_a  + O(h) 
\end{equation}

\begin{equation}
\frac{d\mathbf{v}_a}{dt}  \approx - \frac{P_a}{h \rho_a}\mathbf{E}_{0,a} -  \frac{(\mathbf{V}_a \nabla_{a}) P_a}{\rho_a} + O(h) 
\end{equation}
where $\mathbf{V}$, $\mathbf{R}$ are matrices close to the identity matrix
$\mathbf{I}$ and $\mathbf{E}_{0}$ is a non-vanishing error
vector. $\mathbf{V}$, $\mathbf{R}$ and $\mathbf{E}_{0}$ are all
determined by the particle distribution within smoothing length $h$
(see \cite{Read2009} for detailed definitions).  As
a result, these errors will remain at the same level while we increase
$N$ if $N_{nb}$ is fixed. However, a consistent scheme would
require all the error terms in the above two equations go down as $h
\rightarrow 0$. Similar conclusions can also be drawn for the energy
equation.

Based on the above discussions of the density estimate and the
discretized Euler equations in SPH, the condition $N_{nb} \rightarrow
\infty $, $N \rightarrow \infty $ and $ h \rightarrow 0$ has to be met
in order to have a \textit{consistent and self-consistent}
scheme. Moreover, reducing $h$ will lead to finer time steps 
according to the CFL
condition. If the magnitude of the error from a fixed $N_{nb}$ 
remains
constant, the overall error will build up more
rapidly in a high
resolution run than a lower resolution one.

\section{Error estimate of density estimate}

The error in SPH from the smoothing procedure (independent of particle distribution)
can be approximated to lowest order by:
\begin{equation}
error_s  \propto h^{\alpha}.
\end{equation}
For the usual B-splines \citep{Monaghan2005, Price2012} and the
Wendland functions \citep{Dehnen2012}, this is a second order term
where $\alpha$ = 2. Higher order precision can be achieved by
constructing different classes of smoothing kernels. However, not all
are appropriate
since smoothed estimates of 
positive definite
quantities can take on negative values.

The error from discretizing the integral convolutions in the
smoothing procedure depends on
whether or not particles are 
ordered in space.  Optimistically, if the particles are 
quasi-ordered and not randomly distributed, it has been
conjectured that the 
discretization error can be approximated by \citep{Monaghan1992}
\begin{equation}
 error_d \propto \frac{\log N_{nb}}{N_{nb}}. 
 \end{equation}
This estimate is based on the study of the complexity associated
with the low discrepancy sequence in numerical integration by
\cite{Wozniakowski1991}.  This indicates 
that the actual number of particles
within the volume of each SPH particle should match the
\textit{expected} number of particles. Clearly, this is highly
dependent on the randomness in the actual distribution of SPH
particles.

The pressure force between SPH particles always tends to regulate the
distribution of the SPH neighbors into an isotropic
distribution.  However, this ``restoring" pressure force does not push
the particles into an exact desired distribution since the number of
neighbors is finite.  Hence, the velocity field exhibits noise
on small scales.  Such velocity noise will further induce fluctuations
in the density field.  In realistic applications, especially with highly
turbulent flows, it is questionable whether the pressure forces can
effectively regulate SPH particles into a quasi-ordered
configuration.

On the other hand, if the distribution of particles is \textit{truly}
random, the convergence of SPH should follow (slow) Monte-Carlo
behavior:
\begin{equation}
 error_d \propto \frac{1}{\sqrt{N_{nb}}}.
 \end{equation}

We further parameterize the dependence of discretization error on
the number of neighbors from these two extreme situations as
\begin{equation}
 error_d \propto  N_{nb}^{-\gamma}, 
 \end{equation}
 where $\ 0.5 < \gamma < 1$.

From the perspective of convergence, we require
that as the total number of SPH particles is made larger, we will
be able to simulate finer scale structure.  The smoothing error
is already consistently reduced by making $h$ smaller.
At the same time, we
must also use more neighbors in the discrete sums to combat the discretization
error, which will eventually dominate if no action is taken.  We
can write
these conditions as
\begin{equation}
N \rightarrow \infty ,  h  \rightarrow 0,  N_{nb} \rightarrow \infty.
\end{equation}

If we parameterize the dependence of smoothing length $h$ on $N$ written by
\begin{equation}
 h \propto N^{-1/\beta},
 \end{equation}
then the relation between $N_{nb}$ and $N$ is just
\begin{equation}
N_{nb} \propto \frac{h^3} {V/N} \propto N^{1-3/\beta}.
\end{equation}

The requirements of convergence impose the follow conditions on the dependence of $\beta$ on $N$:
\begin{equation}
\beta > 0,  \ so\  h \rightarrow 0 \ as\  N  \rightarrow \infty, 
\end{equation}
and 
\begin{equation}
 \beta > 3,  \ so\  N_{nb} \rightarrow \infty \  as\  N \rightarrow \infty.
 \end{equation}

%A strong dependence of $h$ on $N_{nb}$ would drastically increase the CPU time for increasing higher resolution. 

For any case $\beta > 3$, there is a power-law dependence of $N_{nb}$
on $N$. In order to balance the smoothing and discretization errors,  i.e. $error_d \sim error_s$, 
we have $ \alpha \sim \gamma (\beta - 3) $.

The dependence of of $h$ and $N_{nb}$ can be expressed as,  \begin{equation}\label{eq:ngbscaling}
\beta \sim \frac{\alpha}{\gamma} + 3,   h \propto N^{-\gamma/(3\gamma + \alpha)},  N_{nb} \propto N^{\alpha/(3\gamma+\alpha)}.
\end{equation}

For commonly used smoothing kernel forms, we have $\alpha = 2$. Recall for random distributions we have $\gamma = 0.5$ and for
quasi-ordered distributions $\gamma = 1$, so $\beta$ for these two
extremes is $\beta = 7$ and $\beta = 5$. The dependence of $N_{nb}$ on $N$ is thus between $[N^{0.4}, N^{0.57}]$.  This suggests a
simple intermediate choice is given by:

\begin{equation}\label{eq:ngbscaling_common}
\beta \sim 6,   h \propto N^{-1/6},  N_{nb} \propto N^{0.5}.
\end{equation}

In detail, other choices would follow if higher-order smoothing
kernels are used, which would require an even stronger 
scaling of $N_{nb}$ with $N$.  This can be readily seen from the expression of $N_{nb}$ 
in equation \ref{eq:ngbscaling}, where the power-law index
$\alpha/(3\gamma+\alpha)$ approaches 1 for sufficiently large $\alpha$ if $\gamma$ 
fixed. 
For example we have $\alpha=4$ for the smoothing kernel functions
constructed by \cite{Monaghan1985},  and then $N_{nb}$ should vary between $[N^{0.57}, N^{0.73}]$.  
This indicates that the regularity in the particle distribution,  which poses a strong limit for numerical interpolation, 
has to be improved to be much better than a quasi-ordered configuration if higher order smoothing kernels are employed.

We note that an earlier suggestion of increasing $N_{nb}$ to reduce the error 
was proposed by \cite{Quinlan1996} from 1-D error analysis of SPH.  On the surface, their analysis
 on the arbitrary spaced particles is similar to our result of quasi-order distribution. The difference here is that
the discretization error in SPH in multi-dimensions is much worse than
 the \textit{truncation error} inferred from the second Euler-MacLaurin formula in 1-D 
 given by \cite{Quinlan1996}.

\section{State of the ART SPH: Wendland function without paring instability and inviscid SPH}

A direct application with the scaling in equation \ref{eq:ngbscaling}
will not work with classical SPH codes.  If more than $\sim 60$
neighbors are used in the discrete estimates, the SPH particles 
will quickly
form close pairs, as demonstrated in \cite{Springel2011} and
\cite{Price2012}. Such a configuration actually makes the
discretization errors worse, which is opposite to our purpose.

%\textcolor{red}{This explains why we use Wendland function to perform the resolution study in this work. Previous SPH kernels cause particles clump to form close pairs within the smoothing kernel when $N_{nb}$ is large ($~\sim$ 60 for the cubic spline). Also this introduce the use of Cullen viscosity to give results closer to the behavior of inviscid fluid with SPH. }

For the kernel functions widely used in SPH codes, the
function $\nabla W$ tends to flatten when two particles are close
together.  Consequently, the net repulsive force also diminishes between
close pairs.  When a large number of neighbors $N_{nb}$ is used,
particles will be less sensitive to small perturbations within the
smoothing kernel. In reality, the density estimate error fluctuates around
the true value as a function of $N_{nb}$ as shown in
\cite{Dehnen2012}.  The whole system thus favors a distribution of
particles with close pairs which minimizes the total energy. Hence the
entire system would not achieve the resolution one aims for. This
clumping instability is the primary reason that the convergence study
with variable $N_{nb}$ is not possible in SPH simulations
\citep{Price2012, Bauer2012, Vogelsberger2012, Hayward2013} with
conventional smoothing kernels.  With the
cubic spline, sometimes additional repulsive forces are added in the
inner smoothing length in the equations of motion to fight
against the pairing instability, as in \cite{Kawata2013}. \cite{Read2009}
instead proposed a centrally peaked kernel to mitigate against clumping. 

\cite{Dehnen2012} have shown that Wendland functions 
are free from the clumping instability for
large $N_{nb}$
when used to
perform smoothing in SPH.
The desirable property of the Wendland functions is
that they are smooth and have non-negative Fourier transforms in a
region with compact support.  Similarly, \cite{GarciaSenz2014} has shown
that the sinc family of kernels is able to avoid particle clumping.
In what follows, we adopt the Wendland
$\rm{C^4}$ function, defined on $[0, 1]$ as   

\begin{equation}
W(q;h) = \frac{495}{32\pi h^3}(1-q)^6(1+6q+\frac{35}{3}q^2), 
\end{equation}
in the P-Gadget-3 code (an updated version of
Gadget-2 \citep{Springel2005}) to study the convergence rate of SPH.

We also update the viscosity switch in P-Gadget-3 using the method by
\cite{Cullen2010} to effectively reduce the artificial viscosity in
shear flows while maintaining a good shock capturing capability.  This
effectively reduces the viscosity in the flows producing results
closer to the \textit{inviscid} case. This switch also greatly
enhances the accuracy of SPH in the subsonic regime especially where
large shear is present. The time integration accuracy is also improved
following the suggestions by \cite{Saitoh2009} and \cite{Durier2012}.

\begin{figure}
\begin{center}
\begin{tabular}{c}
\resizebox{3.2in}{!}{\includegraphics{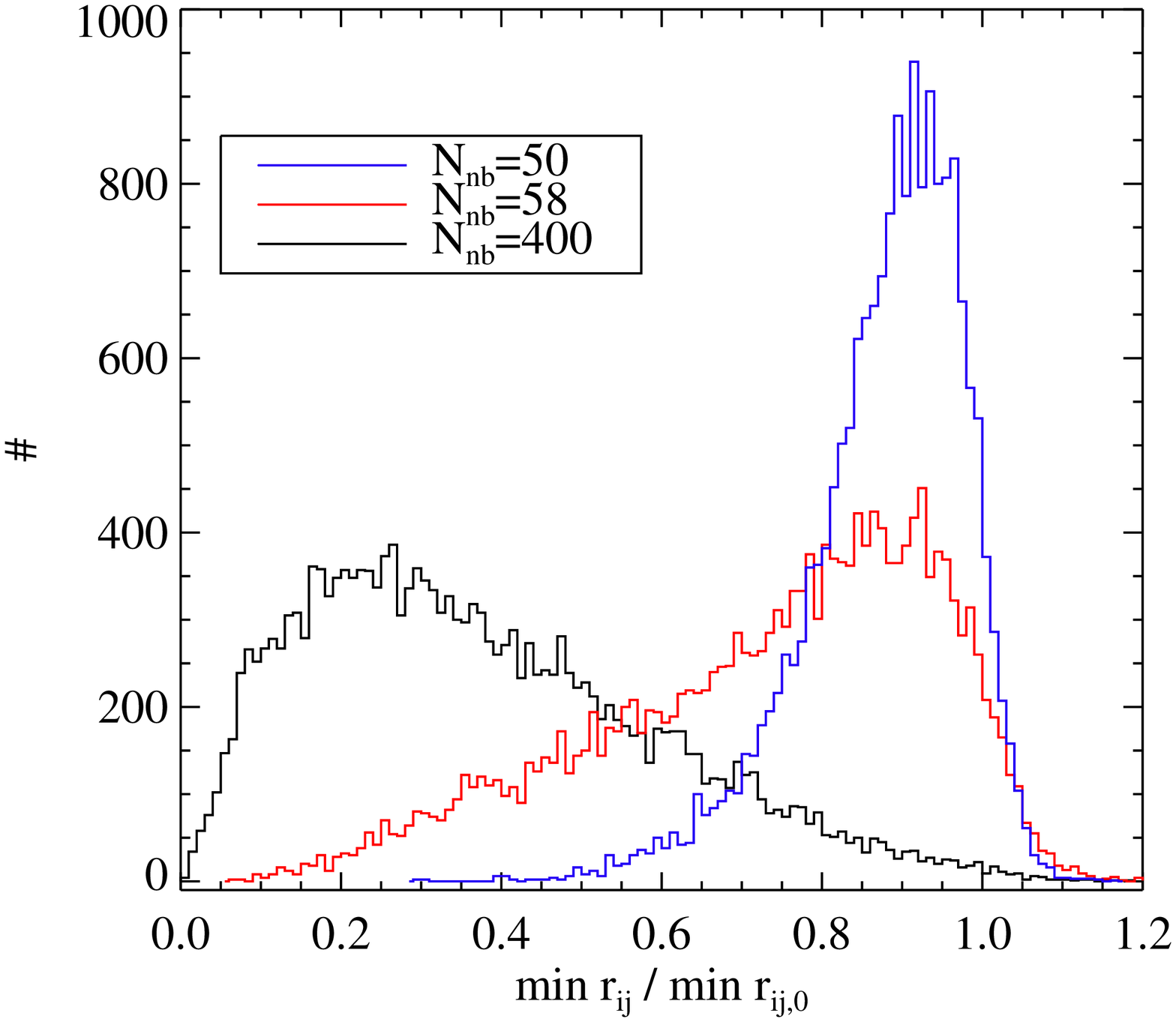}}
\end{tabular}
\begin{tabular}{c}
\resizebox{3.2in}{!}{\includegraphics{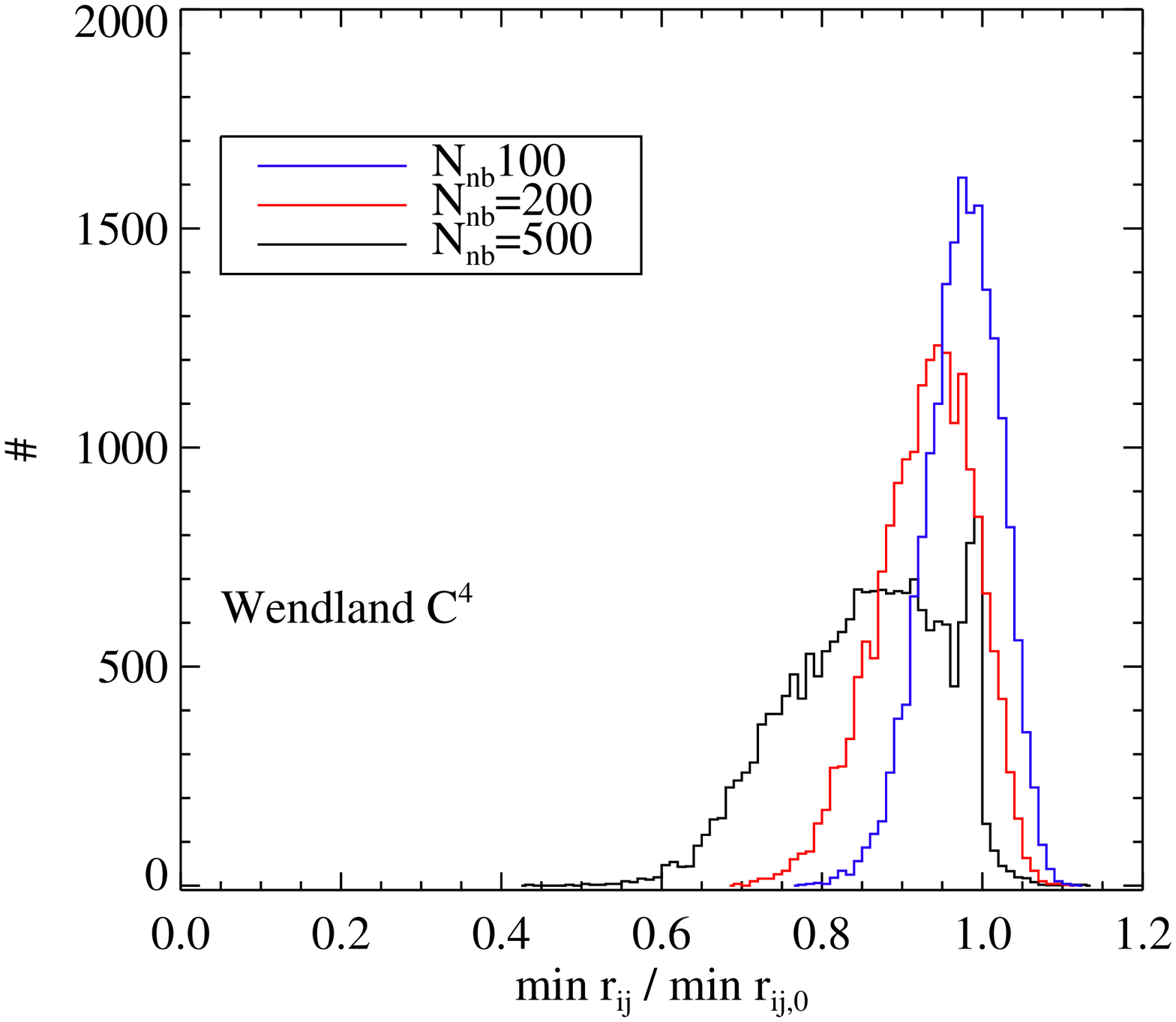}} 
\end{tabular}
\caption{\label{fig:pair_instability}  
\textit{Upper:} The distribution of distances to the nearest
neighbor $\rm{r}_{ij}$ normalized to the initial particle spacing
$\rm{r}_{ij, 0}$ in the Gresho vortex test at $t = 1$, for the cubic
spline in a portion of the simulation domain. \textit{Lower:} The
same histogram for a simulation with the Wendland $\rm{C^4}$
function. In the upper panel, once the number of neighbor exceeds a
threshold, $N_{nb} = 58$ for the cubic spline, close pairs of SPH
particles form quickly.  As a result, the volume associated with
each SPH particle is not well-sampled by their neighbors. This
particle clumping effectively reduces the resolution of SPH.  As a
comparison, this instability is not observed for the simulation with
the Wendland $\rm{C^4}$ function.}
\end{center}
\end{figure}

To verify the absence of the clumping instability with our new code, we
plot the distribution of distances to the closest neighbor particle
from a realistic situation.  In Figure~\ref{fig:pair_instability}, we
compare the performance of the Wendland $\rm{C^4}$ with the original
cubic spline for the Gresho vortex test \citep{Gresho1990}. The details
of the set-up can be found in the following sections. In the upper
panel of Figure~\ref{fig:pair_instability}, we show the distribution
of the distance to the nearest neighbor for each SPH particle in a
subdomain of the simulation at $t = 1$ with the cubic spline. This test
involves strong shearing motions from a constant rotating velocity field.
At $t = 1$, the vortex has completed almost a full rotation. The
distance to the nearest neighbor is further normalized to the mean
interparticle separation in the initial conditions. 

With $N_{nb} = 50$, the distribution of distances shows that most of
the particles are still well-separated from one other. This situation
quickly worsens as soon as we use $N_{nb} = 58$, which is the critical
number for the cubic spline reported by \cite{Dehnen2012}.  A
significant portion of the particles now have their closest neighbor
within a factor of $0.2$ of the initial separation.  For an even higher
$N_{nb} = 400$, most of the particles have formed close pairs as
indicated by the peak near $0$.  As a comparison, the lower panel of
Figure~\ref{fig:pair_instability} shows that close particle pairs are
not present with the Wendland $\rm{C^4}$ as we vary $N_{nb}$ from 100
to 200 (equivalent to $\sim$55 with cubic spline;
see \citealt{Dehnen2012}) to 500. 
This test confirms the good performance of SPH against
the pairing instability with the Wendland function and gives us
confidence that we can use this code to perform further tests to study
the convergence rate with varying $N_{nb}$.

\section{Density Estimate}

\subsection{A set of randomly distributed points}

\begin{figure}[h]
\begin{center}
\begin{tabular}{c}
\resizebox{3.2in}{!}{\includegraphics{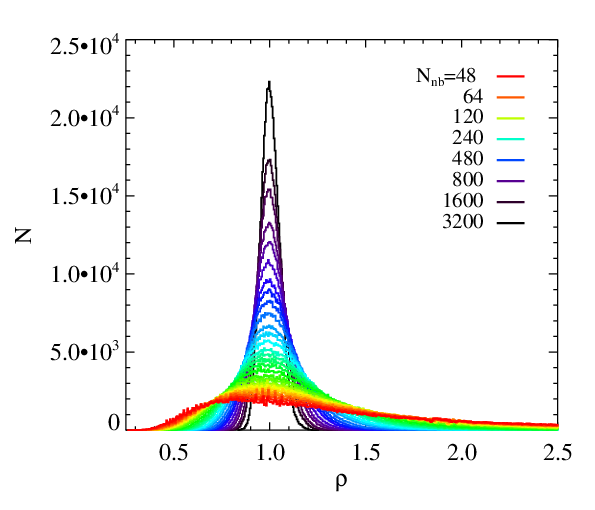}}
\end{tabular}
\begin{tabular}{c}
\resizebox{3.2in}{!}{\includegraphics{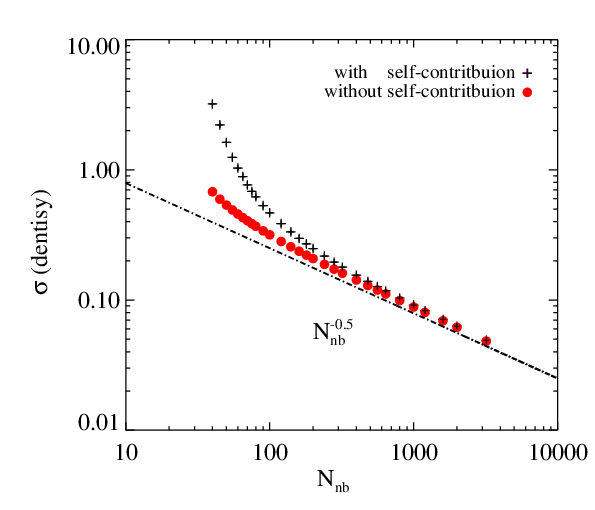}} 
\end{tabular}
\caption{\label{fig:random_data_set} 
\textit{Upper:} The distribution of density field estimates
based on an SPH
smoothing routine with Wendland $C^4$ function on a random set of
points
with varying $N_{nb}$. The color scheme represents different curves
with increasing $N_{nb}$ with the color from red to purple. The
standard deviation decreases as we use a larger $N_{nb}$ in the SPH
density estimate. For $N_{nb} > 1000$, the result is close to a
Gaussian peaked at $\rho = 1$. \textit{Lower:} The standard deviation
measured in the density distribution as a function of varying
$N_{nb}$.  The expected trend of $\sigma \propto N^{-0.5}$ is given by
the dashed-dotted line.  The SPH density estimate with and without
the self-contribution term is indicated by black crosses and red filled
circles.  Self-contribution produces an overestimate of the density where
particles are randomly distributed. }
\end{center}
\end{figure}

We use a randomly distributed particle set to test the density
estimate in SPH.  $64^3$ particles are distributed randomly within a
3-D box with unit length.  We use the density estimate routine in SPH
with varying $N_{nb}$ from 40 to 3200 to derive the density field
accordingly. In Figure~\ref{fig:random_data_set}, we plot the
histogram of the density estimated at the position of each particle
for varying $N_{nb}$. The variance within the density field is largest
for small $N_{nb}$ values, which are shown in the red colors.  As the
number of neighbors increases, the distribution slowly approaches a
Gaussian like distribution peaked at $\rho = 1$.  This behavior shows
the trade-off between variance and bias in the SPH density
estimate. Larger $N_{nb}$ hence larger $h$ will increasingly smear out
fluctuations on short scales.  However this does not show the actual
accuracy of the SPH density estimate.

In fact, the \textit{precise} density field in this set-up is
difficult to give but the behavior of the density field can be seen
from a pure statistical argument. To be more specific, the number of
particles falling within a fixed volume associated with each particle
follows a \textit{Poisson} distribution.  The expected number of such
occurrences goes linearly with larger volume (larger $N_{nb}$). As a
result, the standard deviation of the density distribution follows a
$N_{nb}^{-0.5}$ trend. So should the trend of the standard deviation
of the density distribution given by SPH if the density estimate is
accurate.  However, this is not strictly true as we show in the lower panel of
Figure~\ref{fig:random_data_set}, where the black crosses are the
measured standard deviation of the density distribution with varying
$N_{nb}$. There is a significant spread that is much greater for small
$N_{nb}$ than the guiding line indicating $N_{nb}^{-0.5}$. The
deviation is already significant for $N_{nb} = 200$, which is the
recommended number by \cite{Dehnen2012} based on a shock tube test. In
the upper panel of Figure~\ref{fig:random_data_set}, we can also see
there is a long tail for small $N_{nb}$ values. This suggests there is
an overestimate of density where we suspect the self-contribution of
each particle to the density estimates plays an exaggerated role.

In order to verity this, we carry out density estimates with
self-contribution excluded for the same configuration of
particles. The standard deviation of the density distribution as a
function of $N_{nb}$ is shown in the red filled circles in the lower
panel of Figure~\ref{fig:random_data_set}. This brings the relation
much closer to the expected $N_{nb}^{-0.5}$ line where the deviation
at small $N_{nb}$ is greatly reduced due to the subtraction of
the self-contribution term.

Excluding the self-contribution is not the usual practice in modern
SPH codes. \cite{Flebbe1994} actually advocated to use equation
\ref{eq:sphdensity} with self-contribution excluded to give a better
density calculation based on a similar comparison in 2-D.
\cite{Whitworth1995} however argues that the distribution of SPH
particles is essentially different from a random distribution.  If
particles are distributed randomly, individual particles do not care
about the locations of other particles.  In SPH, due to the repulsive
pressure force, each SPH particle is trying to establish a zone where
other SPH particles cannot easily reside. However, as we see in
Figure~\ref{fig:pair_instability}, SPH forms close pairs so that the
argument by \cite{Whitworth1995} is no longer valid.  Note that it is
not uncommon to see a value above $N_{nb} = 58$ in the literature with
the cubic spline.  The SPH density estimate will then give a biased result
for irregularly distributed particles.

This comparison actually favors the use of a large $N_{nb}$ where the
distribution is highly disordered.  Although the above experiment is
based on a situation where the particle distribution is truly random,
the realistic situation is between such a truly random distribution and
a quasi-regular one.  However, quasi-regular distributions are
hard to achieve in multi-dimensional flows because of shear.

\subsection{A set of points in a glass configuration}

A random distribution of particles, as discussed in the above section,
is an extreme case for SPH simulations.  We relax these particles
according to the procedure described in \cite{White1996} to allow them
to evolve into a glass configuration. Given an $r^{-2}$ repulsive
force, the particles settle in a equilibrium distribution which is
quasi force-free and homogeneous in density. This mimics the other
extreme case where SPH particles are well regulated by pressure
forces.  Nevertheless, the outcome of this relaxation is not perfect
where tiny fluctuations in the density field are still present because
of the discrete nature of the system.

\begin{figure}[h]
\begin{center}
\begin{tabular}{c}
\resizebox{3.2in}{!}{\includegraphics{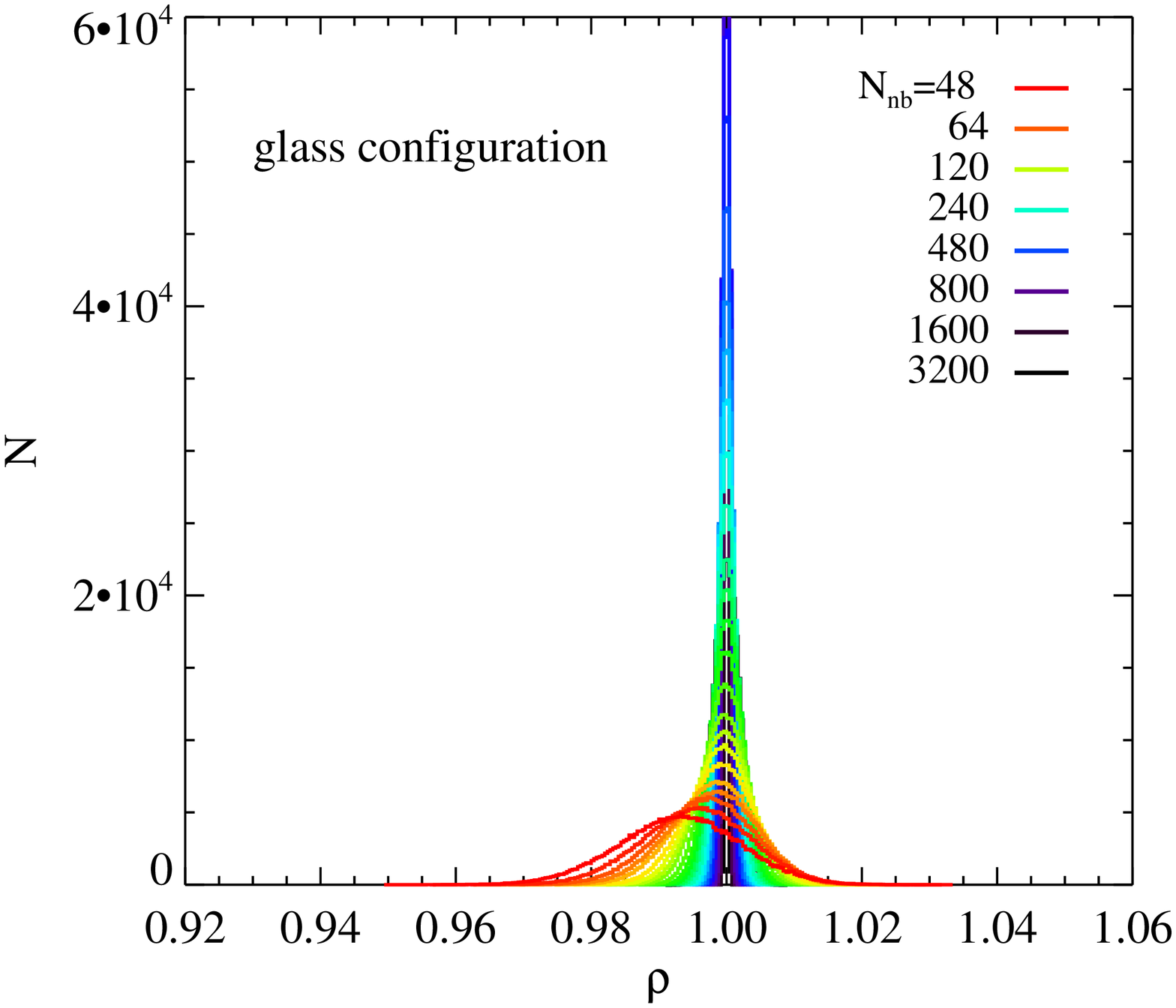}}
\end{tabular}
\begin{tabular}{c}
\resizebox{3.2in}{!}{\includegraphics{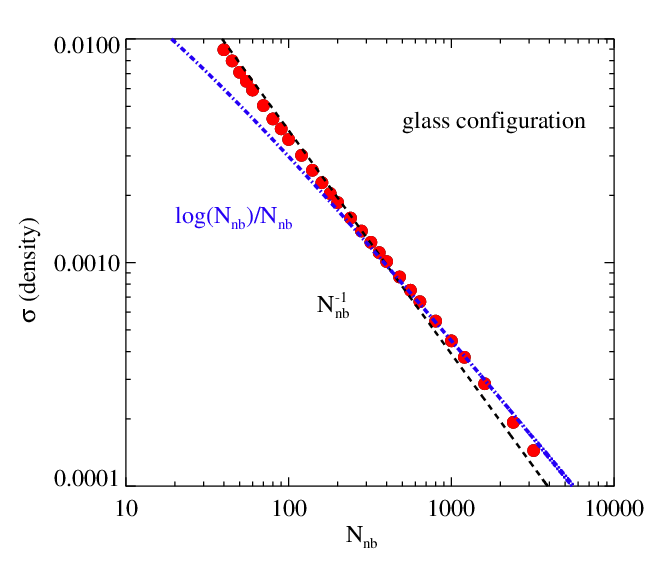}} 
\end{tabular}
\caption{\label{fig:glass_data_set} 
\textit{Upper:} Similar to Figure~\ref{fig:random_data_set}, the
distribution of density field estimates based on an SPH 
smoothing routine with the
Wendland $C^4$ function on a glass configuration data
set for varying $N_{nb}$. The distribution of the density estimates
is significantly narrower than the random point set in
Figure~\ref{fig:random_data_set}. The color schemes represents
different curves with increasing $N_{nb}$ for color 
going from red to
purple. The standard deviation also decreases as we use a larger
$N_{nb}$ in the SPH density estimate but at a much faster rate compared to
Figure~\ref{fig:random_data_set}. \textit{Lower:} The standard
deviation measured in the density distributions as a function of
varying $N_{nb}$.  A trend of $\sigma \propto N^{-1}$ rather than
$\sigma \propto N^{-0.5}$ is seen as indicated by the dashed-dotted
line. }
\end{center}
\end{figure}

We now measure the SPH density estimate on this glass configuration for
different $N_{nb}$ and estimate the standard deviation of the density for
each $N_{nb}$.  As the upper panel in Figure~\ref{fig:glass_data_set}
shows, the density distribution is much narrower than that
for a random set of points.  For
sufficiently high $N_{nb}$, the distribution approaches a
Dirac-$\delta$ distribution. The convergence is also much faster than
in Figure~\ref{fig:random_data_set}. In fact, we observe a $N_{nb}^{-1}$
trend for the glass configuration. For this set up, there is no need
to exclude the self-contribution term, which is consistent with
\cite{Whitworth1995}.

This measured $N_{nb}^{-1}$ rate is quite interesting in
itself. \cite{Monaghan1992} conjectured that the discretization error
with SPH behaves as $\log(N_{nb})/N_{nb}$.  This means that the randomness
in the distribution of SPH particles is closer to a low
discrepancy sequence rather than to a truly random one. Our
experiment shows this $N_{nb}^{-1}$ (neglecting the $\log(N_{nb})$)
rate is certainly appropriate for a glass configuration, which is
carefully relaxed to an energy minimum state. However, this rate may
not hold for a general applications.

Thus, these two examples support the dependence of the density
estimate error on $N_{nb}$ as in the previous analysis.  Namely,
$N_{nb}^{-0.5}$ for truly random data points and $N_{nb}^{-1}$ for a
distribution with good order. Realistic applications will fall between
these two extremes and could be more biased towards the
$N_{nb}^{-0.5}$ case.

\subsection{Errors in the volume estimate}

\begin{figure}[h]
\begin{center}
\begin{tabular}{c}
\resizebox{3.2in}{!}{\includegraphics{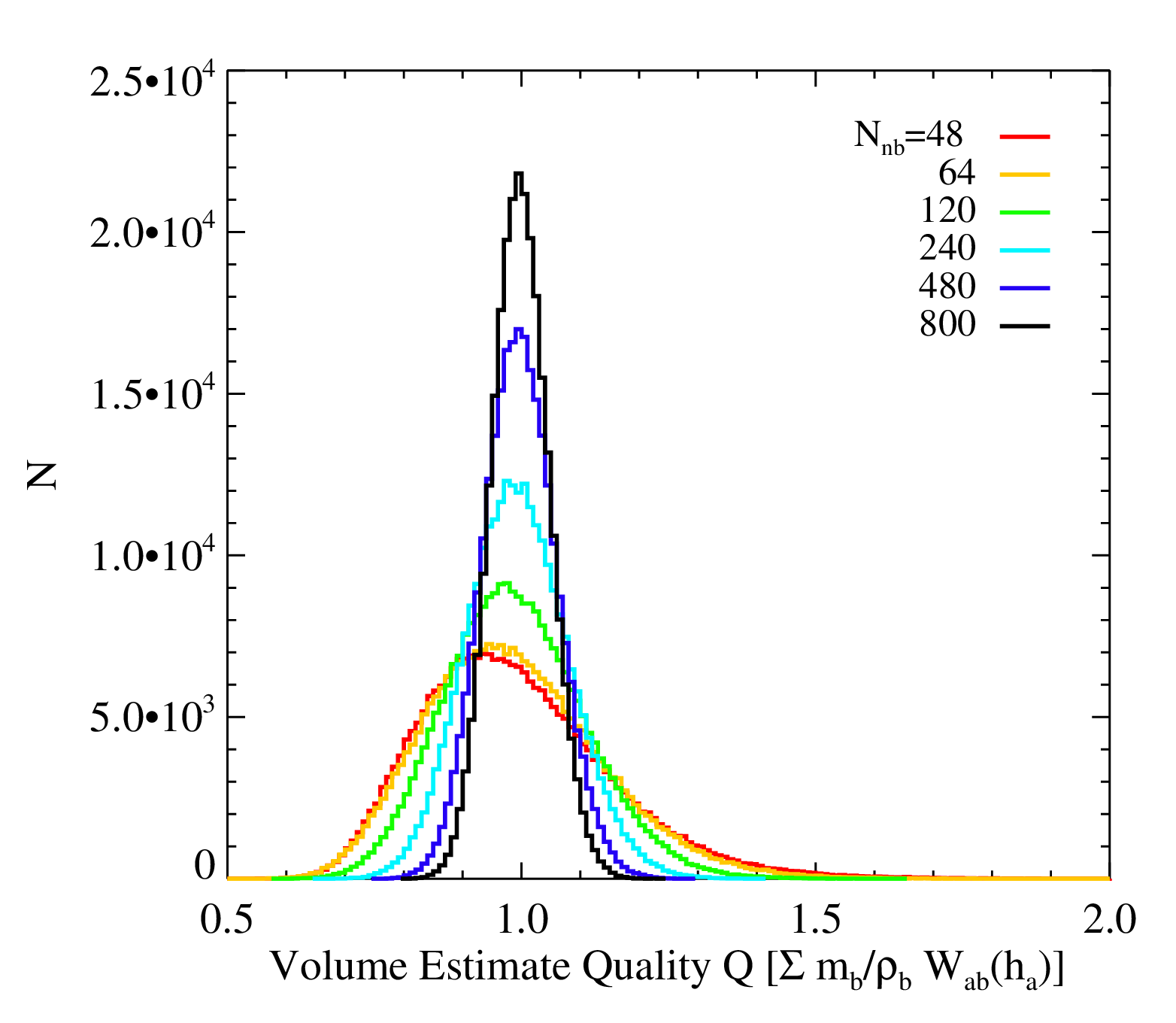}}
\end{tabular}
\begin{tabular}{c}
\resizebox{3.2in}{!}{\includegraphics{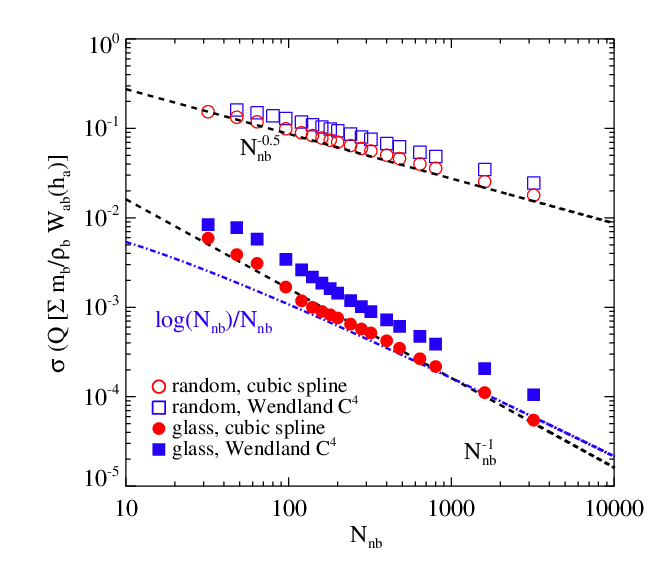}}
\end{tabular}
\caption{\label{fig:volume_error} . 
\textit{Upper}: The distribution of the volume estimate quality parameter
$Q_a = \sum m_b/ \rho_bW_{ab}(h_a)$ for the random point set used in
Figure~\ref{fig:random_data_set} for different $N_{nb}$.  This
quantifies the deviation from an exact partition of unity. The
distribution of $Q$ slowly approaches a narrower normal distribution
centered at $Q = 1$ with increasing $N_{nb}$.  This indicates that the
volume estimate error will eventually decrease to $0$ as we have
$N_{nb} \rightarrow \infty$ when the distribution of Q is essentially
a Dirac-$\delta$ function at 1. \textit{Lower}: The relation between
$Q$ and $N_{nb}$ measured with Wendland $C^4$ function and with a
cubic spline for the random point set and for the glass configuration
respectively.  For the glass configuration, the distribution of $Q$
converges to a Dirac-$\delta$ function at a rate of $N^{-1}$ for both
of the smoothing functions. However, the convergence rate goes as
$N^{-0.5}$ for both of the smoothing functions with the random point
set.}
\end{center}
\end{figure}

We have noted that a constant scalar field is not exactly
represented by the SPH method. Based on our discussion of the
self-consistency of the SPH density estimation, we can further measure the
magnitude of this inconsistency and its dependence on the number of
neighbors $N_{nb}$.  Define a parameter $Q$ as 
\begin{equation}
Q_a =
\sum_{j}^{N_{nb}} \frac{m_b}{\rho_b} W_{ab}(h_a),
\end{equation}
 which represents
the SPH estimate of a constant scalar field $A = 1$. This parameter
$Q$ should follow a peaked distribution around 1 with some errors.  We
compute this parameter with a simple extra routine after the density
calculation in the code since this parameter requires the SPH density
field as input. In Figure~\ref{fig:volume_error}, we plot the
distributions of the calculated values of $Q$ for randomly
distributed particles with several different $N_{nb}$ values. The
distributions of $Q$ are indeed rather broad around
1. Similarly as the errors in the density estimate, the standard
deviation of the distribution is consistently larger for smaller
$N_{nb}$ values. Some extreme values of Q with small $N_{nb}$ are even
on the order of unity. The peaks of the distribution with small
$N_{nb}$ are actually slightly below 1, which indicates some bias
towards a density overestimate. This is consistent with the above
discussion of the self-contribution to the density overestimate at
small $N_{nb}$ values.

For this random distribution of particles, increasing $N_{nb}$ will
lead to a more accurate volume estimate by reducing the inconsistency
in the density estimate.  For example, the spread of the distribution
of $Q$ for $N_{nb} = 800$ is greatly reduced compared to $N_{nb} =
48$. As we repeat this experiment with even more neighbors, the
distribution of $Q$ approaches a Dirac-$\delta$ function
eventually with the error in the volume estimate reduced to zero.  In
this limit, the self-consistency of SPH is finally restored.  However,
as in the lower panel of Figure~\ref{fig:volume_error}, such an
improvement depends slowly on the number of number $N_{nb}$ at
a rate $\sigma (Q) \propto N_{nb}^{-0.5}$. 

We also consider the cubic
spline in this test, which is included in the lower panel of
Figure~\ref{fig:volume_error}. The dependence of Q on $N_{nb}$ for a
cubic spline follows the same $N_{nb}^{-0.5}$ trend. The cubic spline
shows a slightly lower magnitude for the same $N_{nb}$ compared with
the Wendland $C^4$ function. This is because the Wendland $C^4$ is slightly
more centrally peaked than the cubic spline. To have the same
effective resolution length, a higher number of $N_{nb}$ is required
for Wendland $C^4$ (see the table of such scaling in
\cite{Dehnen2012}). The behavior of $Q$ with respect of $N_{nb}$ is
consistent with the two smoothing functions.

We then repeat this calculation with the particles in a glass
configuration. The results of the Wendland $C^4$ function and cubic
splines are presented with the solid symbols in
Figure~\ref{fig:volume_error}.  As we see from the previous section,
the relaxed particle distribution reduces the errors
in the density estimate. The inconsistency in the volume estimate, as
measured in $Q$, declines as a function of $\sigma (Q) \propto
N_{nb}^{-1}$ in this case. This is true for both the Wendland $C^4$
function and cubic spline.

In practice, when we use SPH, the true distribution is unknown. So the
error in the density estimate is difficult to quantify.  We suggest a
simple procedure to measure Q in simulations to give us an estimate
of the overall quality of the density estimate. Based on the
comparison between the $Q$ value for randomly distributed particles
and particles in a glass configuration, a typical number $N_{nb} = 32
\ \rm{or} \ 64$ with cubic spline as often adopted in the literature
leads to a volume estimate error between 1$\%$ and 10$\%$ from the
true value.  However, a glass configuration is a rather idealized and
we suspect that SPH particles will not maintain such good order in
most applications.

This error in the volume estimate, which is self-inconsistency in the
SPH density estimate, also has dynamical consequences as the volume
estimate, 
\begin{equation}
d\mathbf{r} \approx \frac{m}{\rho}, 
\end{equation}
is used for the derivation of the equations of motion. In addition
to the error terms derived by \cite{Read2009}, an error on the order
of a percent level is already present from the density estimate step
itself. The quality of the particle distribution is crucial to
minimizing this error. However, once the particles deviate from a
regular distribution, the errors in the equations of motion will pose
another challenge for the particles to return back to an ordered
configuration. One way to impose a small error in the volume estimate
is to use a large $N_{nb}$. Increasing $N_{nb}$ can regulate this
error with a dependence between $N_{nb}^{-0.5}$ and $N_{nb}^{-1}$.

\section{Dynamical tests}

In this section, we perform dynamical tests with SPH for several
problems to study the convergence rates in each situation.  The
evolution in multi-dimensional flows is more complex than in 1-D and,
as noted in \S 1, one-dimensional tests cannot be used to 
judge convergence of SPH in multi-dimensions.
The
test problems are chosen so analytic behaviors are known in
advance (\cite{Springel2010, Read2012, Owen2014, Hu2014}).  Also the test problems consider both smooth flows and highly
turbulent flows. These test problems are purely \textit{hydro} as no
further compilations from other factors such as gravity are
present.  We also do not consider problems with shocks for our
tests as the presence of discontinues will reduce the code to 
being at best first order accurate.

\subsection{Gresho Vortex problem}

The first problem is the Gresho Vortex test \citep{Gresho1990}. We
quantify the $L_1$ velocity error with SPH and its convergence rate
as a function of the total number of SPH particles. This test involves a
differentially rotating vortex with uniform density in centrifugal
balance with pressure and azimuth velocity specified according to
\begin{equation}
%\[
    P(r)= 
\begin{cases}
    P_{0} + 12.5r^2 &  (0 \leq r < 0.2)\\
    P_{0} + 12.5r^2 + 4 \\ - 20r + 4\ln(5r) &  (0.2 \leq r < 0.4)\\    
    P_{0} + 2(2\ln2 - 1)              & (r \geq 0.4)
\end{cases}
%\]
\end{equation}

and

%\[
\begin{equation}
    v_{\phi}(r)=
\begin{cases}
    5r &  (0 \leq r < 0.2)\\
    2 -5r & (0.2 \leq r < 0.4)\\    
    0              & (r \geq 0.4).
\end{cases}
\end{equation}
%\]

The value of $P_{0}$ is taken to be $P_0 = 5$ following
\cite{Springel2010} and
\cite{Read2012}, as in the original paper of
\cite{Gresho1990}. \cite{Dehnen2012} and
\cite{Hu2014} also investigate the
behavior by varying this constant background pressure $P_0$. The
velocity noise will decline for decreased $P_0$ such that the system
is more stable. This test problem is suitable for examining the noise
in the velocity field since the density field is uniform. The vortex
should be time independent having the initial analytical
structure. 

\cite{Springel2010} however found that this test poses a
challenge to SPH since the vortex quickly breaks up with
Gadget. \cite{Dehnen2012},
\cite{Read2012},
and \cite{Hu2014}  have tested their SPH
formulation with the same problem and found improvements with a better
parameterization of artificial viscosity. The convergence rate
reported by \cite{Dehnen2012} and
\cite{Hu2014} is similar to the one given by
\cite{Springel2010}, $N^{-0.7}$, while \cite{Read2012} give a rate of
$N^{-1.4}$.  This discrepancy is due to a modified version of the
equations of motion by \cite{Read2012}.  The set-up of this problem
has three sharp transition points which is problematic for SPH. This
test is also very sensitive to the particle noise which will can induce
a false triggering of artificial viscosity.

\begin{figure}[h]
\begin{center}
\includegraphics[scale=0.40]{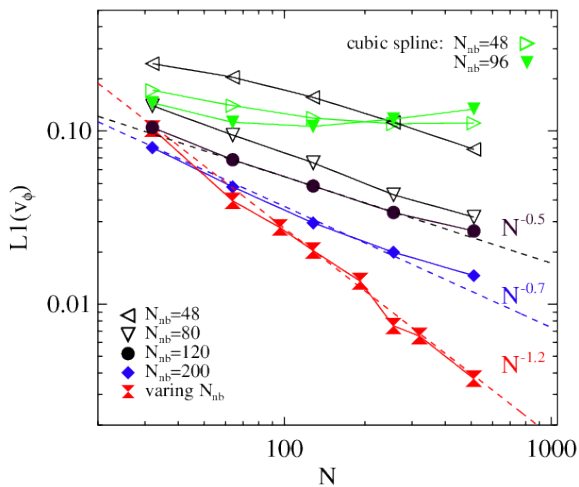}
\caption{\label{fig:gresho_convergence} .  
Convergence rate of the velocity field in the Gresho vortex test as a
function of $N_{1D}$ with SPH. Blue filled circles are the
results with a fixed number of neighbors $N_{nb} = 200$ while the black
filled squares are the ones with a smaller number of neighbors $N_{nb}
= 120$. The convergence rate shows a $N^{-0.7}$ behavior for the
former and $N^{-0.5}$ for the latter. The red squares are with
increased $N_{nb}$ as a function of total number of SPH particles
$N$.  The convergence rate for this run scales almost as
$N^{-1.2}$. Such a convergence rate is close to $N^{-1.4}$ for the
grid-based code in \cite{Springel2010arepo}. }
\end{center}
\end{figure}

We initialize the particles on an N$\times$N$\times$16 lattice. The
number of particles vertically assures that we can extend $N_{nb}$ to
several thousand for each particle without overlap.  The $L_1$
velocity error as a function of the number of effective 1-D resolution
elements $N$ is shown in Figure~\ref{fig:gresho_convergence}.  The
black squares are the simulations with $N_{nb} = 120$ and the blue
filled circles are the ones with $N_{nb} = 200$. For the red diamonds,
we run the simulation with increasing $N_{nb}$ as $N_{nb} \propto
N^{3*0.4}$ starting from $N_{nb} = 120$ at the lowest resolution $N =
32$. The convergence rates are as $N^{-0.5}$, $N^{-0.7}$ and
$N^{-1.2}$ respectively.  The former two are consistent with
\cite{Springel2010}, \cite{Dehnen2012} and \cite{Hu2014}.  Also, a
larger number of neighbors gives a slightly faster convergence rate in
this test ($N^{-0.5}$ vs. $N^{-0.7}$). For $N_{nb} = 120$ and $N_{nb}
= 200$, the $L_{1}$ velocity error at the highest resolution at
$N=500$ is actually \textit{above} the convergence rate.  This is
what is expected based on the argument for consistency in the 
SPH method. A fixed
$N_{nb}$ may work well for a range of different resolutions until the
resolution is ultimately so high that the errors from the density
estimate, from the discretized continuity equation and momentum
equation, are not consistently reduced according to the corresponding
resolution length $h$.

The result with varying $N_{nb}$, on the other hand, shows a much
faster convergence rate as $N^{-1.2}$ with the highest resolution
at $N = 320$ with $N_{nb} = 1900$. This validates the arguments
presented earlier and shows the result given by SPH does
\textit{converge} to the proper solution and the convergence rate is
improved in this test problem as we increase $N_{nb}$ when $h
\rightarrow 0$. Such a convergence rate is actually close to the rate $p
= 1.4$ measured with moving mesh code Arepo \citep{Springel2010arepo}
for the same problem.

We have also included two series of the Gresho vortex test with the
cubic spline as in the green symbols in
Figure~\ref{fig:gresho_convergence}. The performance with the cubic
spline is disappointing in this test. With $N_{nb} = 48$, the $L_1$
error reaches to a minimum value at an intermediate resolution at $N =
128$. For $N_{nb} = 96$, where the clumping instability is reducing
the effective resolution, the minimum $L_1$ error is achieved at $N =
64$. In terms of the absolute error, the highest resolution actually
fails to give the best results. This behavior is also observed in the
Gresho vortex test and the KH instability test in \cite{Springel2010}.

\subsection{Isentropic Vortex problem}

%\textcolor{red}{This quantifies the L2 error in the Yee Vortex with SPH and its convergence rate. This shows a second order accuracy can be achieved provided that enough $N_{nb}$ is used so the discretization error will not degrade the overall quality of the simulation. If such requirement is not met, the simulation will give inconsistent result. }

This is another vortex problem described by \cite{Yee2000} and
employed by \cite{Calder2002} and \cite{Springel2011}. It also has a
time-invariant analytic solution. Unlike the previous test, the
density and velocity profiles are both smooth. This problem will test
the convergence rate of the solver for smooth flows.  We generate the
initial conditions in a box of size $[-5,5]^2$ with periodic boundary
conditions. Similar to the previous test, several additional identical
layers are stacked along the $z$ direction to get a thin slab
configuration. The velocity field is set up as

\begin{equation}
    v_{x}(x,y) =  -y \frac{\beta}{2\pi} \exp(\frac{1-r^2}{2}), 
\end{equation}
\begin{equation}
    v_{y}(x,y) =  x \frac{\beta}{2\pi} \exp(\frac{1-r^2}{2}). 
    \end{equation}

\begin{figure}[h]
\begin{center}
\includegraphics[scale=0.40]{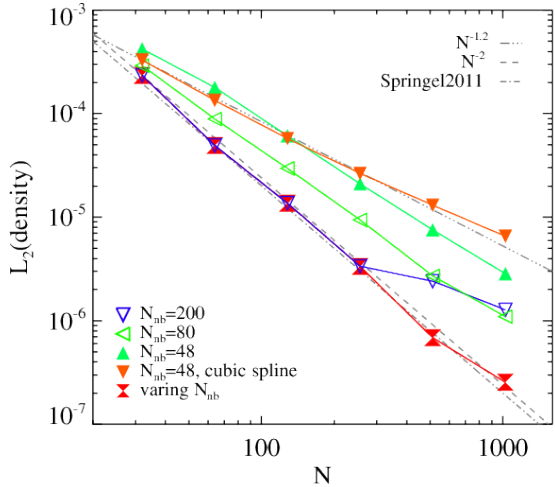}
\caption{\label{fig:yee_convergence} .  
Convergence rate of the isentropic vortex test as a function of
$N_{1D}$ with SPH code. The error is computed from the $L_2$ norm of
the density field at $t = 8.0$ when the vortex has completed a full
rotation.  Blue filled circles are the result for a fixed
number of
neighbors $N_{nb} = 200$. The convergence rate exhibits a $N^{-2}$
behavior
except for the two high resolution runs. The red filled diamonds are
the two high resolution runs for increasing $N_{nb}$ and now
the error is
consistent with the expected $N^{-2}$ trend. }
\end{center}
\end{figure}

The distribution of the density field and the internal energy per unit
mass are calculated according to the following temperature
distribution

\begin{equation}
 T(x, y) \equiv P/\rho = T_{\infty} - \frac{(\gamma - 1)\beta^2}{8 \gamma \pi^2}\exp(1-r^2),  
\end{equation}
as $\rho = T^{1/(\gamma - 1)}$ and $u = T/(\gamma -1)$.  Consequently,
the entropy $P/\rho^{\gamma}$ is a constant within the domain. As in
\cite{Springel2011}, we then measure the $L_2$ error in the density field
at $t = 8.0$ obtained with our code with respect to the analytical
density field for different resolutions.  The result if shown in
Figure~\ref{fig:yee_convergence} where the blue filled circles are
obtained with $N_{nb} = 200$ while the red diamonds are with
varying $N_{nb}$.  The dashed line is a $N^{-2}$ trend with the
dashed-dotted line taken from \cite{Springel2011} measured with the
Arepo code.

The overall magnitude of the error with SPH is also close to the one
for the Arepo simulation in \cite{Springel2011}. In fact, the result
for $N_{nb} = 200$ with SPH shows a second order convergence for the
first four resolutions, but for the two highest resolution runs the
$L_2$ density error starts to reach a plateau.  Beginning with $N_{nb}
= 200$ at $N = 256$, we re-simulate the two high resolution runs with
$N_{nb} \propto N^{3*0.5}$. As a result, the overall error is reduced
and is now consistent with the expected rate of $N^{-2}$.

The above result based on the isentropic vortex problem shows that SPH
\textit{can} be a second-order accurate scheme as long as the other
errors do not dominate over the ``smoothing error" introduced from the
density estimate. Again, the condition $N\rightarrow \infty , h
\rightarrow 0 \ $ and $ \ N_{nb} \rightarrow \infty$ needs to be
satisfied. As for the fixed $N_{nb}$ run, starting from $N=512$ the
other errors rather than the ``smoothing error" become dominant and
start to degrade the performance of SPH.

In order to see the effects of the discretization error, we have
repeated this experiment with smaller $N_{nb}$ and also using
the cubic spline. For the two series of $N_{nb} = 80$ and $N_{nb}
= 48$, as in green symbols in Figure~\ref{fig:yee_convergence}, the
convergence rate systematically slows down from $N^{-1.7}$ to
$N^{-1.3}$ from a second order convergence rate. The convergence rate
with the cubic spine, where $N_{nb} = 48$, is the slowest
one.  Moreover, it shows signs of turning over at the highest
resolution with the cubic spline.

\subsection{Subsonic turbulence}

%\textcolor{red}{This demonstrate that increased $N_{nb}$ will deliver better results in the energy power spectrum such that more extended inertia range and reduced velocity noise at small scale will obtained in a driven subsonic turbulence test.}

In this section, we move to a more complex problem involving the
chaotic feature of multi-dimensional flows with SPH. In particular, we
simulate subsonic turbulence in the gas (which is weakly
compressible). Following \cite{Bauer2012} and
\cite{Price2012turbulence}, we employ a periodic box with unit length
filled with an isothermal gas $\gamma = 1$ with unit mean gas density and
unit sound speed. The initial positions of the particles are
set on a cubic lattice. The stirring force field is set up in Fourier
space and we only inject energy between k = 6.28 and k = 12.56. The
amplitude of these modes follows a -5/3 law as: $P(k) \propto k^{-5/3}$,
while the phases are calculated for an Ornstein$-$Uhlenbeck process
$$x_t = f x_{t-\Delta t} + \sigma \sqrt{(1-f^2)}Z_n,$$
such that a smoothly varying driving field with finite correlation
time is obtained.  The parameters in the Ornstein$-$Uhlenbeck process
are identical to the ones adopted by \cite{Bauer2012} and
\cite{Price2012turbulence} as well as \cite{Hopkins2013}. A Helmholtz
decomposition in k-space is applied before the Fourier transformation
to yield a pure solenoidal component in the driving force field. At
every time step, before the hydro force calculation, this driving
routine is called to compute the acceleration. The magnitude of the
driving force is adjusted such that the $r.m.s.$ velocity is at $\sim
0.2 c_s$ once the system reaches a quasi-equilibrium state. Besides
the routine to calculate the driving force, we have also included a
function to measure the velocity power spectrum on the fly. The
turbulent velocity power spectrum is calculated whenever a new
snapshot is written. We use the ``nearest neighbor sampling" method
for our estimate. Finally the 1D power spectrum is obtained by
averaging the velocity power with each component in fixed
$|\mathbf{k}|$ bins. We use the time-averaged velocity power spectra
between $t = 10$ and $t = 20$ for the following comparison.

\begin{figure}[h]
\begin{center}
\includegraphics[scale=0.50]{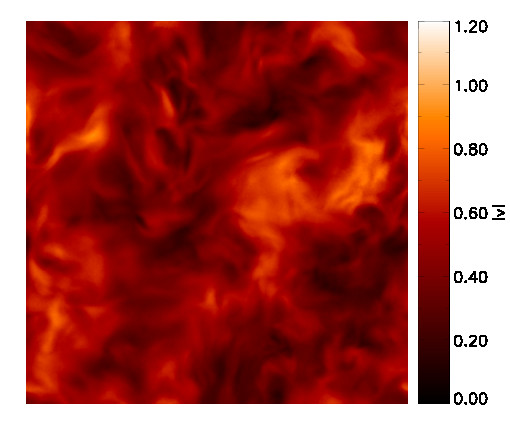}
\caption{\label{fig:subsonic_high_res} .  
A high resolution subsonic turbulence test with $256^3$ particles showing
the magnitude of the velocity field at $t = 10.0$ for a thin slice in
the box. A cross comparison between this figure and the ones in
\cite{Bauer2012} and \cite{Price2012} at the same resolution shows
that our SPH code is able to resolve the finest structures among all
three. The velocity field is qualitatively closer to the one
simulated with a grid based code. This is a result of the use of the
artificial viscosity switch proposed by \cite{Cullen2010} and the Wendland $C^4$
smoothing kernel.  $N_{nb}= 200$ is used in this example.}
\end{center}
\end{figure}

We emphasize that our SPH code is able to give improved
performance in this test over the previous studies of \cite{Bauer2012}
and \cite{Price2012}. A snapshot of the velocity field is shown in
Figure~\ref{fig:subsonic_high_res} in a high resolution test with
$256^3$ particles. Our SPH code is able to resolve finer
structures compared to the other two studies noted. The velocity field
is qualitatively closer to the one simulated with grid based
codes. This is a result of the use of the artificial viscosity switch
proposed by \cite{Cullen2010} and the Wendland $C^4$ smoothing kernel.

\begin{figure}[h]
\begin{center}
\begin{tabular}{ll}
\resizebox{1.7in}{!}{\includegraphics{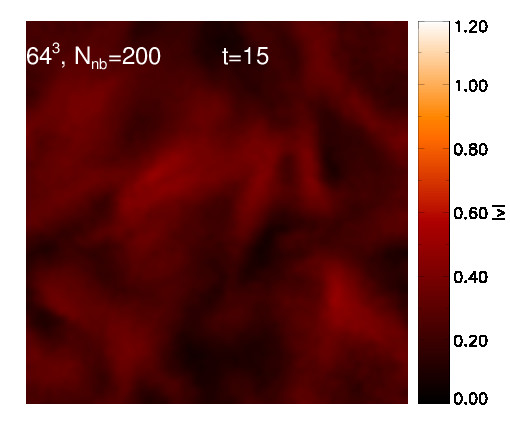}}
\resizebox{1.7in}{!}{\includegraphics{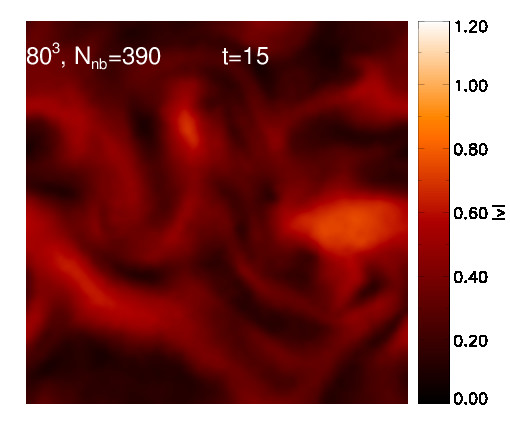}}
\end{tabular}
\begin{tabular}{ll}
\resizebox{1.7in}{!}{\includegraphics{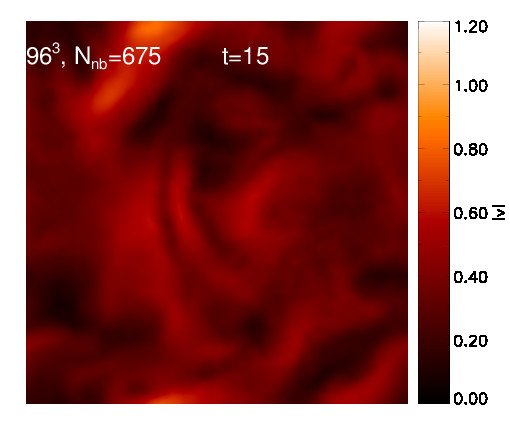}}
\resizebox{1.7in}{!}{\includegraphics{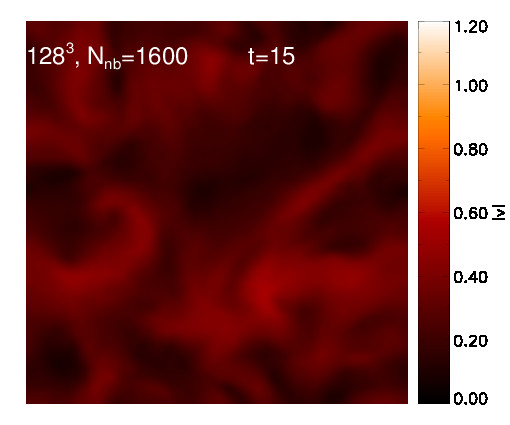}}
\end{tabular}
\caption{\label{fig:turbulence_resolution} 
Projected maps of velocity field for four subsonic turbulence tests. The
total number of SPH particles is increased from $64^3$ to $128^3$
while the number of number of neighbors $N_{nb}$ is also increased
accordingly in order to have the same smoothing length $h$ for all the
simulations. }
\end{center}
\end{figure}

Four different resolutions, $64^3$, $80^3$, $96^3$ and $128^3$, are
used for a resolution study.  We actually fix the smoothing length $h$
for all the simulations based on $N_{nb} = 200$ for the $64^3$
resolution run and subsequently scale $N_{nb}$ for the other three
cases. This is done to directly test the effect of varying $N_{nb}$ on
the hydrodynamics by controlling the same resolution length $h$ so
that the smoothing error is fixed. The projected velocity maps of
these four resolutions are shown in
Figure~\ref{fig:turbulence_resolution}. The velocity fields are similar
to one another between these four simulations.

\begin{figure}[h]
\begin{center}
\includegraphics[scale=0.4]{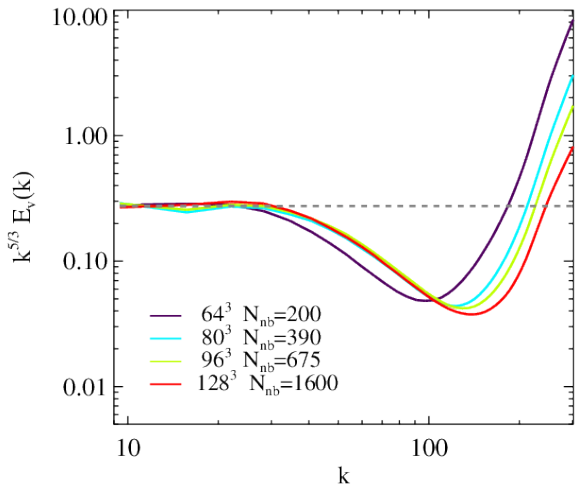}
\caption{\label{fig:turbulence_convergence} 
The velocity power spectrum in a driven subsonic turbulence test at
different resolutions. The velocity power is multiplied by $k^{5/3}$,
where $k$ is the wave number, such that the inertial range will show
up as a horizontal line. Different total numbers of SPH particles are
used while the number of neighbors are set up such that the
``smoothing length" $h$ is the same for all the simulations.  Even
with the same nominal ``h-resolution", the $128^3$ run shows
an improvement over the $64^3$ run, especially for the location of
the point of
turn-over from the minimum on small scales and the increased
velocity power on large scales. }
\end{center}
\end{figure}

In Figure~\ref{fig:turbulence_convergence}, we compare the velocity
power spectrum for these four resolutions.  We multiply the power
spectrum by $k^{5/3}$ so that the inertial range described by $P(k)
\propto k^{-5/3}$ shows up as a horizontal line. The power spectrum
shows good agreement with the $k^{-5/3}$ law on large scales. The
velocity power goes down significantly below that once the wave number
$k$ is above 40. However, there is significant power contained on the
smallest scales. The spatial extent where neighbor particles are
searched corresponds to $k \sim 110$, which is the range where the
turn-over is observed. This turn-over can be seen as the result of
dissipation of velocity noise on the resolution scale \citep{Hopkins2013}.

Improvements with varying $N_{nb}$ in this test problems can be
seen. The effect of varying $N_{nb}$ on the velocity power spectrum is
visible on both large and small scales. With more neighbors, the
velocity field on the kernel scale (smoothing length $h$) is better
regulated as the power is reduced by an order of magnitude. This can
be also seen in that the turn-over point for higher resolution
increasingly shifts to a higher $k$ value. On large scales where the
modes are well resolved, more neighbors also help to bring the power
spectrum into better agreement with the $k^{-5/3}$ law. This can be
verified by comparing the red and purple lines for $128^3$ and $64^3$
in Figure~\ref{fig:turbulence_convergence}. The variance on the power
spectrum between $k = 10$ and $k = 30$ is reduced for the latter
case. The velocity power contained within $k = 30$ and $k \sim 110$ is
also significantly higher in the $128^3$ run with $N_{nb} = 1600$ than
$64^3$ with $N_{nb} = 200$. The result is that there is increasingly
less power in the noisy motion on the kernel scale but more power in a
coherent fashion. The conclusion is that $N_{nb}$ does play an
important role in this subsonic turbulence test as suggested by
\cite{Bauer2012}. This also agrees with the consistency requirement
for SPH in order to recover the continuous limit that the
``discretization errors" also need to be reduced with $h \rightarrow
0$.  

\begin{figure}[h]
\begin{center}
\includegraphics[scale=0.4]{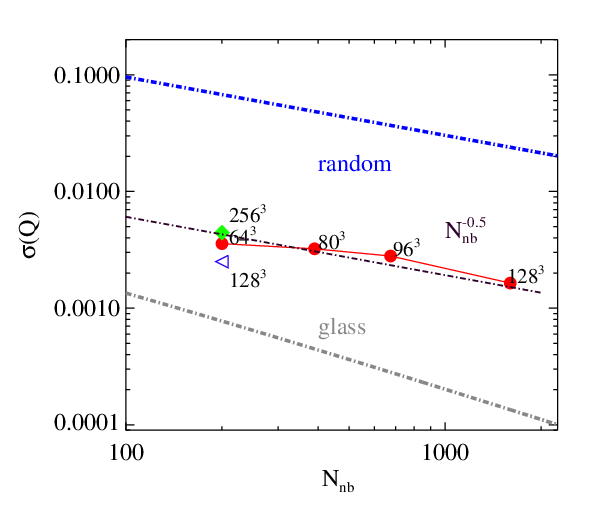}
\caption{\label{fig:volume_error_turbulence} 
The distribution of $Q = \sum m_b/ \rho_b Wij(h_a)$ for several
snapshots at the same $t$ in the subsonic turbulence tests in order to
access the degree of disorder in realistic SPH applications.  The
trend of Q from the static random particles and the glass
configuration are also over-plotted here. The distribution of $Q$ for
these six snapshots is between the two static cases. The trend of $Q$
with respect to $N_{nb}$, however, is quite shallow in these dynamical
cases. These points roughly follow a $N_{nb}^{-0.5}$ trend. SPH is
not able to reorder the particles to a distribution with the degree of
order characterized by a glass configuration. }
\end{center}
\end{figure}

Figure~\ref{fig:volume_error_turbulence} shows the
value of $Q$, as defined above, in the turbulence tests
at $t = 15$.  We have also over-plotted the trend
of Q from the static distribution of random particles and the
glass configuration in the previous section.  This allows us to use
$Q$, which describes the deviation from an exact partition of unity,
to access the degree of disorder in realistic applications of
SPH.  In fact, the distribution of $Q$ for these six snapshots are
between the two static cases. The trend of $Q$ with respect to
$N_{nb}$, however, is quite shallow among these dynamical cases,
indicating a $N_{nb}^{-0.5}$ trend.  SPH is not able to re-order the
particles to a distribution with the degree of order characterized
by the glass configuration. The degree of disorder could be much
higher in post-shock flows frequently encountered in astrophysical
simulations than what we have found in these weakly compressible
turbulence flows with SPH since the density variation is much higher
in the former. 

\section{Conclusions}

The main findings of our study are as follows:

\begin{itemize}

  \item We argue that the requirements of consistency and
    self-consistency within the SPH framework require that
    $N\rightarrow \infty , h \rightarrow 0$ and $N_{nb} \rightarrow
    \infty$ need to be satisfied to obtain the true
    continuum behavior of a flow. We summarize the errors from
    the smoothing step and from the discretization step in the SPH density
    estimate and propose a power-law scaling between the
    desired $N_{nb}$ given the total number of SPH particles $N$ in
    order to have a consistent and convergent scheme. The dependence
    $N_{nb} \propto [N^{0.40}, N^{0.57} ]$ is derived based on a
    balance between these two types of errors for commonly used
    smoothing kernels.

  \item We verify the error dependence on $N_{nb}$ for the
    discretization error with a quasi-ordered glass configuration and
    a truly random one. The range for the power-law index
    between $N_{nb}$ and $N$ is obtained based on these two extreme
    conditions. The discretization error in the density estimate decreases
    roughly as $\log({N_{nb}})/N_{nb}$ for the glass configuration and as
    $N_{nb}^{-0.5}$ for the random distribution.
  
  \item We propose a simple method to calculate a parameter $Q$, which
    measures the deviation from an exact partition of unity in
    SPH. The error in the volume estimate in SPH is further examined
    for a glass configuration and a random one. In
    agreement with the behavior of the density estimate error, this
    quantity also shows the same dependence on $N_{nb}$ in these two
    situations. In realistic applications, the error introduced with
    the density estimate step itself is on the order of several percent for
    the $N_{nb}$ value used in most published work.
  
  \item We use the Wendland $C^4$ function as a new smoothing kernel
    to perform dynamical tests with a varying $N_{nb}$ to avoid the
    clumping instability. We confirm that particle pairs do not form
    even with a large value of $N_{nb}$. The artificial viscosity by
    \cite{Cullen2010} also greatly improves the performance of our SPH
    code. As a consequence, the results from our study are not
    influenced by the numerical artifacts from these two aspects as in
    some of the previous studies.
 
   \item A fixed $N_{nb}$, which indicates an inconsistent scheme,
     gives a slow convergence rate as reported by previous studies in
     the two vortex problems. As shown in the figures, the
     convergence rate levels off and eventually turns over for
     sufficiently fine resolution. Also the error obtained with the
     cubic spline is larger than for the Wendland $C^4$
     function. The cubic spline actually exaggerates the discretization
     error as the Wendland $C^4$ function is much smoother.
   
   \item For smooth flows, as seen in the isentropic vortex problem,
     varying $N_{nb}$ according to the proposed power-law dependence can
     give second-order convergence. The same improvement is
     seen in the Gresho vortex test, where several discontinuities
     are included. The convergence rate $p = 1.2$ with varying
     $N_{nb}$ is closer to the rate $p = 1.4$ measured with the
     second-order accurate moving mesh Arepo code for the same problem.
   
   \item For highly turbulent flows, the velocity noise poses another
     challenge in addition to the errors in the density estimate for the
     SPH particles to retain an ordered configuration. Though our
     code is able to give a better result in the subsonic
     turbulence test compared with previous studies, the velocity
     noise on the kernel scale is always present. This velocity noise
     can be reduced by an order of magnitude with a larger $N_{nb}$,
     but the inertial range resolved with SPH code only shows a slight
     improvement.

    \item We measure the randomness of SPH particles in the above
      subsonic turbulence tests using the calculation of $Q$. This
      further verifies our assumptions when deriving the dependence of
      $N_{nb}$ on $N$. The distribution of SPH particles in this
      example is indeed
      between a truly random configuration and a quasi-ordered case. 
      This approach
      can be used in general to quantify the randomness in SPH in
      applications. Unfortunately, SPH is not able to rearrange the
      particles into a highly ordered configuration characterized by a
      low discrepancy sequence.
  
    \item The usual practice of using a fixed $N_{nb}$ indicates that the
      same level of noise will be present even if higher resolution is
      used. This error can be stated as a ``zeroth order" term which
      is \textit{independent of resolution}. Moreover, the errors
      introduced can be roughly approximated by a Gaussian
      distribution around the true value provided $N_{nb}$ is large
      enough so that the self-contribution term is not significant even for
      highly random distributions.
 
   \end{itemize}

We emphasize that the inconsistency and the errors associated with particle disorder is not a specific feature to a particular SPH code but is rather a 
generic problem. In fact, the set of equations in Gadget are fully conservative since they are derived from the Euler-Lagrange equations (see \citealt{Springel2002}, \citealt{Springel2010} and \citealt{Price2012}). To our knowledge, this approach has not been fully adopted by the community yet. The conservation of mass, linear momentum, angular momentum and total energy has been long recognized to be the unique advantage over other techniques. Indeed, its conservative nature, its robustness and its simple form have made it a popular tool for astrophysical modeling and in other disciplines. However, the errors associated with the finite summation $error_d$, usually termed as ``noise" in the SPH literature, and its impact on the convergence of SPH has not been well understood so far. As we discussed earlier, such an error naturally and inevitably emerges from the density estimate itself and also from the momentum equation when evolving the system.

Obviously, one promising direction to improve the particle method is to use a more accurate volume estimate. It is possible to derive a numerical scheme which can reproduce a polynomial up to the $n^{th}$ order polynomial exactly following the procedure by \cite{Liu2006}. In practice, it is sufficient to eliminate the zero-th order error by renormalization such that the corrected scheme is second-order accurate by symmetry.

Recently, \cite{Hopkins2014b} proposed a new class of gridless Lagrangian methods which use finite-element Godunov schemes. These gridless methods dramatically reduce the low-order errors and numerical viscosity in SPH method and significantly improve accuracy and convergence. They show better performance in resolving fluid-mixing instabilities, shocks and subsonic turbulence, which is critical in a wide range of physical and astrophysical problems. 

We caution against equating numerical precision with physical accuracy of a simulation because both physics and numerics play roles in the modeling of a complex system. For example, we compared cosmological hydrodynamical simulations of a Milky-Way like galaxy using both the improved Gadget code (presented in this paper) and the gridless GIZMO code by \cite{Hopkins2014b} with the same initial conditions, physics prescriptions and resolutions, and found that both simulations produced similar galaxy properties such as disk mass, size and kinematics, although GIZMO resolves the spiral structures better. These results suggest that the robustness of astrophysical simulations depends not only on the numerical algorithms, but also on the physical processes, as highlighted also in a number of recent papers on galaxy simulations (e.g., \citealt{Scannapieco2012, Hopkins2014a, Vogelsberger2013, Vogelsberger2014a, Vogelsberger2014b, Schaye2014}).

\section*{Acknowledgements}
We thank Volker Springel, Phil Hopkins, Mark Vogelsberger, Jim Stone, Andreas Bauer and Daniel Price for valuable discussions on this work. We further thank Paul Torrey for his contributions to the experimental setups in the early stage and suggestions throughout the project. LH acknowledges support from NASA grant NNX12AC67G and NSF grant AST-1312095, YL acknowledges support from NSF grants AST-0965694, AST-1009867 and AST-1412719. We acknowledge the Institute For CyberScience at the Pennsylvania State University for providing computational resources and services that have contributed to the research results reported in this paper. The Institute for Gravitation and the Cosmos is supported by the Eberly College of Science and the Office of the Senior Vice President for Research at the Pennsylvania State University.


\begin{thebibliography}{49}
\expandafter\ifx\csname natexlab\endcsname\relax\def\natexlab#1{#1}\fi

\bibitem[{{Agertz} {et~al.}(2007){Agertz}, {Moore}, {Stadel}, {Potter},
  {Miniati}, {Read}, {Mayer}, {Gawryszczak}, {Kravtsov}, {Nordlund}, {Pearce},
  {Quilis}, {Rudd}, {Springel}, {Stone}, {Tasker}, {Teyssier}, {Wadsley}, \&
  {Walder}}]{Agertz2007}
{Agertz}, O., {Moore}, B., {Stadel}, J., {Potter}, D., {Miniati}, F., {Read},
  J., {Mayer}, L., {Gawryszczak}, A., {Kravtsov}, A., {Nordlund}, {\AA}.,
  {Pearce}, F., {Quilis}, V., {Rudd}, D., {Springel}, V., {Stone}, J.,
  {Tasker}, E., {Teyssier}, R., {Wadsley}, J., \& {Walder}, R. 2007, \mnras,
  380, 963

\bibitem[{{Bauer} \& {Springel}(2012)}]{Bauer2012}
{Bauer}, A., \& {Springel}, V. 2012, \mnras, 423, 2558

\bibitem[{{Calder} {et~al.}(2002){Calder}, {Fryxell}, {Plewa}, {Rosner},
  {Dursi}, {Weirs}, {Dupont}, {Robey}, {Kane}, {Remington}, {Drake}, {Dimonte},
  {Zingale}, {Timmes}, {Olson}, {Ricker}, {MacNeice}, \& {Tufo}}]{Calder2002}
{Calder}, A.~C., {Fryxell}, B., {Plewa}, T., {Rosner}, R., {Dursi}, L.~J.,
  {Weirs}, V.~G., {Dupont}, T., {Robey}, H.~F., {Kane}, J.~O., {Remington},
  B.~A., {Drake}, R.~P., {Dimonte}, G., {Zingale}, M., {Timmes}, F.~X.,
  {Olson}, K., {Ricker}, P., {MacNeice}, P., \& {Tufo}, H.~M. 2002, \apjs, 143,
  201

\bibitem[{{Cullen} \& {Dehnen}(2010)}]{Cullen2010}
{Cullen}, L., \& {Dehnen}, W. 2010, \mnras, 408, 669

\bibitem[{{Dehnen} \& {Aly}(2012)}]{Dehnen2012}
{Dehnen}, W., \& {Aly}, H. 2012, \mnras, 425, 1068

\bibitem[{{Durier} \& {Dalla Vecchia}(2012)}]{Durier2012}
{Durier}, F., \& {Dalla Vecchia}, C. 2012, \mnras, 419, 465

\bibitem[{{Flebbe} {et~al.}(1994){Flebbe}, {Muenzel}, {Herold}, {Riffert}, \&
  {Ruder}}]{Flebbe1994}
{Flebbe}, O., {Muenzel}, S., {Herold}, H., {Riffert}, H., \& {Ruder}, H. 1994,
  \apj, 431, 754

\bibitem[{{Garc{\'{\i}}a-Senz} {et~al.}(2014){Garc{\'{\i}}a-Senz},
  {Cabez{\'o}n}, {Escart{\'{\i}}n}, \& {Ebinger}}]{GarciaSenz2014}
{Garc{\'{\i}}a-Senz}, D., {Cabez{\'o}n}, R.~M., {Escart{\'{\i}}n}, J.~A., \&
  {Ebinger}, K. 2014, \aap, 570, A14

\bibitem[{{Genel} {et~al.}(2013){Genel}, {Vogelsberger}, {Nelson}, {Sijacki},
  {Springel}, \& {Hernquist}}]{Genel2013}
{Genel}, S., {Vogelsberger}, M., {Nelson}, D., {Sijacki}, D., {Springel}, V.,
  \& {Hernquist}, L. 2013, \mnras, 435, 1426

\bibitem[{{Gingold} \& {Monaghan}(1977)}]{Monaghan1977}
{Gingold}, R.~A., \& {Monaghan}, J.~J. 1977, \mnras, 181, 375

\bibitem[{Gresho \& Chan(1990)}]{Gresho1990}
Gresho, P.~M., \& Chan, S.~T. 1990, International Journal for Numerical Methods
  in Fluids, 11, 621

\bibitem[{{Hayward} {et~al.}(2014){Hayward}, {Torrey}, {Springel}, {Hernquist},
  \& {Vogelsberger}}]{Hayward2013}
{Hayward}, C.~C., {Torrey}, P., {Springel}, V., {Hernquist}, L., \&
  {Vogelsberger}, M. 2014, \mnras, 442, 1992

\bibitem[{{Hernquist}(1993)}]{Hernquist1993}
{Hernquist}, L. 1993, \apj, 404, 717

\bibitem[{{Hernquist} \& {Katz}(1989)}]{Hernquist1989}
{Hernquist}, L., \& {Katz}, N. 1989, \apjs, 70, 419

\bibitem[{{Hopkins}(2013)}]{Hopkins2013}
{Hopkins}, P.~F. 2013, \mnras, 428, 2840

\bibitem[{{Hopkins}(2014)}]{Hopkins2014b}
---. 2014, ArXiv e-prints:1409.7395

\bibitem[{{Hopkins} {et~al.}(2014){Hopkins}, {Kere{\v s}}, {O{\~n}orbe},
  {Faucher-Gigu{\`e}re}, {Quataert}, {Murray}, \& {Bullock}}]{Hopkins2014a}
{Hopkins}, P.~F., {Kere{\v s}}, D., {O{\~n}orbe}, J., {Faucher-Gigu{\`e}re},
  C.-A., {Quataert}, E., {Murray}, N., \& {Bullock}, J.~S. 2014, \mnras, 445,
  581

\bibitem[{{Hu} {et~al.}(2014){Hu}, {Naab}, {Walch}, {Moster}, \&
  {Oser}}]{Hu2014}
{Hu}, C.-Y., {Naab}, T., {Walch}, S., {Moster}, B.~P., \& {Oser}, L. 2014,
  \mnras, 443, 1173

\bibitem[{{Kawata} {et~al.}(2013){Kawata}, {Okamoto}, {Gibson}, {Barnes}, \&
  {Cen}}]{Kawata2013}
{Kawata}, D., {Okamoto}, T., {Gibson}, B.~K., {Barnes}, D.~J., \& {Cen}, R.
  2013, \mnras, 428, 1968

\bibitem[{Liu \& Liu(2006)}]{Liu2006}
Liu, M.~B., \& Liu, G.~R. 2006, Appl. Numer. Math., 56, 19

\bibitem[{{Lucy}(1977)}]{Lucy1977}
{Lucy}, L.~B. 1977, \aj, 82, 1013

\bibitem[{Monaghan(1985)}]{Monaghan1985}
Monaghan, J. 1985, Journal of Computational Physics, 60, 253

\bibitem[{{Monaghan}(1992)}]{Monaghan1992}
{Monaghan}, J.~J. 1992, \araa, 30, 543

\bibitem[{{Monaghan}(2005)}]{Monaghan2005}
---. 2005, Reports on Progress in Physics, 68, 1703

\bibitem[{Owen(2014)}]{Owen2014}
Owen, M. 2014, International Journal for Numerical Methods in Fluids, 75, 749

\bibitem[{{Price}(2008)}]{Price2008}
{Price}, D.~J. 2008, Journal of Computational Physics, 227, 10040

\bibitem[{{Price}(2012{\natexlab{a}})}]{Price2012turbulence}
---. 2012{\natexlab{a}}, \mnras, 420, L33

\bibitem[{{Price}(2012{\natexlab{b}})}]{Price2012}
---. 2012{\natexlab{b}}, Journal of Computational Physics, 231, 759

\bibitem[{Quinlan {et~al.}(2006)Quinlan, Basa, \& Lastiwka}]{Quinlan1996}
Quinlan, N.~J., Basa, M., \& Lastiwka, M. 2006, International Journal for
  Numerical Methods in Engineering, 66, 2064

\bibitem[{{Read} \& {Hayfield}(2012)}]{Read2012}
{Read}, J.~I., \& {Hayfield}, T. 2012, \mnras, 422, 3037

\bibitem[{{Read} {et~al.}(2010){Read}, {Hayfield}, \& {Agertz}}]{Read2009}
{Read}, J.~I., {Hayfield}, T., \& {Agertz}, O. 2010, \mnras, 405, 1513

\bibitem[{Robinson \& Monaghan(2012)}]{Robinson2012}
Robinson, M., \& Monaghan, J.~J. 2012, International Journal for Numerical
  Methods in Fluids, 70, 37

\bibitem[{{Saitoh} \& {Makino}(2009)}]{Saitoh2009}
{Saitoh}, T.~R., \& {Makino}, J. 2009, \apjl, 697, L99

\bibitem[{{Saitoh} \& {Makino}(2013)}]{Saitoh2013}
---. 2013, \apj, 768, 44

\bibitem[{{Scannapieco} {et~al.}(2012){Scannapieco}, {Wadepuhl}, {Parry},
  {Navarro}, {Jenkins}, {Springel}, {Teyssier}, {Carlson}, {Couchman}, {Crain},
  {Dalla Vecchia}, {Frenk}, {Kobayashi}, {Monaco}, {Murante}, {Okamoto},
  {Quinn}, {Schaye}, {Stinson}, {Theuns}, {Wadsley}, {White}, \&
  {Woods}}]{Scannapieco2012}
{Scannapieco}, C., {Wadepuhl}, M., {Parry}, O.~H., {Navarro}, J.~F., {Jenkins},
  A., {Springel}, V., {Teyssier}, R., {Carlson}, E., {Couchman}, H.~M.~P.,
  {Crain}, R.~A., {Dalla Vecchia}, C., {Frenk}, C.~S., {Kobayashi}, C.,
  {Monaco}, P., {Murante}, G., {Okamoto}, T., {Quinn}, T., {Schaye}, J.,
  {Stinson}, G.~S., {Theuns}, T., {Wadsley}, J., {White}, S.~D.~M., \& {Woods},
  R. 2012, \mnras, 423, 1726

\bibitem[{{Schaye} {et~al.}(2014){Schaye}, {Crain}, {Bower}, {Furlong},
  {Schaller}, {Theuns}, {Dalla Vecchia}, {Frenk}, {McCarthy}, {Helly},
  {Jenkins}, {Rosas-Guevara}, {White}, {Baes}, {Booth}, {Camps}, {Navarro},
  {Qu}, {Rahmati}, {Sawala}, {Thomas}, \& {Trayford}}]{Schaye2014}
{Schaye}, J., {Crain}, R.~A., {Bower}, R.~G., {Furlong}, M., {Schaller}, M.,
  {Theuns}, T., {Dalla Vecchia}, C., {Frenk}, C.~S., {McCarthy}, I.~G.,
  {Helly}, J.~C., {Jenkins}, A., {Rosas-Guevara}, Y.~M., {White}, S.~D.~M.,
  {Baes}, M., {Booth}, C.~M., {Camps}, P., {Navarro}, J.~F., {Qu}, Y.,
  {Rahmati}, A., {Sawala}, T., {Thomas}, P.~A., \& {Trayford}, J. 2014, ArXiv
  e-prints:1407.7040

\bibitem[{{Schuessler} \& {Schmitt}(1981)}]{Schussler1981}
{Schuessler}, I., \& {Schmitt}, D. 1981, \aap, 97, 373

\bibitem[{{Springel}(2005)}]{Springel2005}
{Springel}, V. 2005, \mnras, 364, 1105

\bibitem[{{Springel}(2010{\natexlab{a}})}]{Springel2010arepo}
---. 2010{\natexlab{a}}, \mnras, 401, 791

\bibitem[{{Springel}(2010{\natexlab{b}})}]{Springel2010}
---. 2010{\natexlab{b}}, \araa, 48, 391

\bibitem[{{Springel}(2011)}]{Springel2011}
---. 2011, ArXiv e-prints:1109.2218

\bibitem[{{Springel} \& {Hernquist}(2002)}]{Springel2002}
{Springel}, V., \& {Hernquist}, L. 2002, \mnras, 333, 649

\bibitem[{{Vogelsberger} {et~al.}(2013){Vogelsberger}, {Genel}, {Sijacki},
  {Torrey}, {Springel}, \& {Hernquist}}]{Vogelsberger2013}
{Vogelsberger}, M., {Genel}, S., {Sijacki}, D., {Torrey}, P., {Springel}, V.,
  \& {Hernquist}, L. 2013, \mnras, 436, 3031

\bibitem[{{Vogelsberger} {et~al.}(2014{\natexlab{a}}){Vogelsberger}, {Genel},
  {Springel}, {Torrey}, {Sijacki}, {Xu}, {Snyder}, {Bird}, {Nelson}, \&
  {Hernquist}}]{Vogelsberger2014a}
{Vogelsberger}, M., {Genel}, S., {Springel}, V., {Torrey}, P., {Sijacki}, D.,
  {Xu}, D., {Snyder}, G., {Bird}, S., {Nelson}, D., \& {Hernquist}, L.
  2014{\natexlab{a}}, \nat, 509, 177

\bibitem[{{Vogelsberger} {et~al.}(2014{\natexlab{b}}){Vogelsberger}, {Genel},
  {Springel}, {Torrey}, {Sijacki}, {Xu}, {Snyder}, {Nelson}, \&
  {Hernquist}}]{Vogelsberger2014b}
{Vogelsberger}, M., {Genel}, S., {Springel}, V., {Torrey}, P., {Sijacki}, D.,
  {Xu}, D., {Snyder}, G., {Nelson}, D., \& {Hernquist}, L. 2014{\natexlab{b}},
  \mnras, 444, 1518

\bibitem[{{Vogelsberger} {et~al.}(2012){Vogelsberger}, {Sijacki}, {Kere{\v s}},
  {Springel}, \& {Hernquist}}]{Vogelsberger2012}
{Vogelsberger}, M., {Sijacki}, D., {Kere{\v s}}, D., {Springel}, V., \&
  {Hernquist}, L. 2012, \mnras, 425, 3024

\bibitem[{{White}(1996)}]{White1996}
{White}, S.~D.~M. 1996, in Cosmology and Large Scale Structure, ed.
  R.~{Schaeffer}, J.~{Silk}, M.~{Spiro}, \& J.~{Zinn-Justin}, 349

\bibitem[{{Whitworth} {et~al.}(1995){Whitworth}, {Bhattal}, {Turner}, \&
  {Watkins}}]{Whitworth1995}
{Whitworth}, A.~P., {Bhattal}, A.~S., {Turner}, J.~A., \& {Watkins}, S.~J.
  1995, \aap, 301, 929

\bibitem[{Wozniakowski(1991)}]{Wozniakowski1991}
Wozniakowski, H. 1991, Bulletin (New Series) of the American Mathematical
  Society, 24, 185

\bibitem[{Yee {et~al.}(2000)Yee, Vinokur, \& Djomehri}]{Yee2000}
Yee, H., Vinokur, M., \& Djomehri, M. 2000, Journal of Computational Physics,
  162, 33

\end{thebibliography}
\end{document}